\documentclass[lettersize,journal]{IEEEtran}
\usepackage{amsmath,amsfonts}
\usepackage{algorithmic}
\usepackage{algorithm}
\usepackage{array}
\usepackage[caption=false,font=normalsize,labelfont=sf,textfont=sf]{subfig}
\usepackage{textcomp}
\usepackage{stfloats}
\usepackage{url}
\usepackage{verbatim}
\usepackage{graphicx}
\usepackage{cite}
\usepackage{pifont}
\usepackage{hyperref}
\usepackage{wrapfig}
\usepackage{array}
\usepackage{xcolor,colortbl} 
\usepackage{arydshln} 

\graphicspath{{figs/}{figures/}{pictures/}{images/}{./}} 

\newcommand{\paragraphHeading}[1]{\vspace{4px}\noindent\textbf{#1}}
\newcommand{\static}{\emph{static}}
\newcommand{\animated}{\emph{animated}}
\newcommand{\immersive}{\emph{immersive}}

\makeatletter
\def\subsubsection{\@startsection{subsubsection}
                                 {3}
                                 {\z@}
                                 {0ex plus 0.1ex minus 0.1ex}
                                 {0ex}
                                 {\normalfont\normalsize\itshape}}
\makeatother

\newcommand{\dpA}{Cinematic camera shots and transitions}
\newcommand{\dpB}{Intuitive encoding between phenomenon and representation}
\newcommand{\dpC}{Smoothly and naturally moving element}
\newcommand{\dpD}{Direct manipulation of camera and visualization}
\newcommand{\dpE}{Realistic appearance}
\newcommand{\dpF}{Dynamic dimension}

\begin{document}

\title{Understanding the Impact of Spatial Immersion\\ in Web Data Stories}
\author{Seon Gyeom Kim, Juhyeong Park, Yutaek Song, Donggun Lee, Yubin Lee, \\ Ryan Rossi, Jane Hoffswell, Eunyee Koh, Tak Yeon Lee
\thanks{Seon Gyeom Kim, Juhyeong Park, Yutaek Song, Donggun Lee, Yubin Lee and Tak Yeon Lee are with the KAIST, Korea}
\thanks{Ryan Rossi, Jane Hoffswell, Eunyee Koh are with Adobe Research, USA}
}

\markboth{Journal of \LaTeX\ Class Files,~Vol.~14, No.~8, August~2021}%
{Shell \MakeLowercase{\textit{et al.}}: A Sample Article Using IEEEtran.cls for IEEE Journals}


\maketitle

\begin{abstract}
An increasing number of web articles engage the reader with the feeling of being immersed in the data space. However, the exact characteristics of spatial immersion in the context of visual storytelling remain vague. For example, what are the common design patterns of data stories with spatial immersion? How do they affect the reader's experience? To gain a deeper understanding of the subject, we collected 23 distinct data stories with spatial immersion, and identified six design patterns, such as cinematic camera shots and transitions, intuitive data representations, realism, naturally moving elements, direct manipulation of camera or visualization, and dynamic dimension. Subsequently, we designed four data stories and conducted a crowdsourced user study comparing three design variations (static, animated, and immersive). Our results suggest that data stories with the design patterns for spatial immersion are more interesting and persuasive than static or animated ones, but no single condition was deemed more understandable or trustworthy.
\end{abstract}

\begin{IEEEkeywords}
Data and knowledge visualization, Human information processing
\end{IEEEkeywords}

\section{Introduction}
What is immersion? Pioneers in the domain have commonly used the metaphor of being surrounded by a completely different reality that takes over user perception \cite{ermi2007analyzing, mcmahan2013immersion, ryan2015narrative, Witmer1998, murray1997hamlet}. However, the term immersion continues to be applied inconsistently in many fields of research, including virtual environments \cite{slater1997framework}, video games \cite{mcmahan2013immersion}, film studies \cite{ROONEY2012405}, and interactive media \cite{ryan2015narrative}. Nilsson et al. \cite{nilsson2016immersion} summarized them into two opposing views: (1) immersion as system technology and (2) immersion as user experience. Those who follow the first view might consider immersion as an objective measurable property of the system or the device surrounding the user, such as 3D-capable wall displays or head-mounted displays~(HMD) for virtual and augmented reality~(VR, AR) \cite{slater2003}. In contrast, proponents of the second view, including us in this paper, would argue that immersion can be studied as a psychological state characterized by perceiving oneself enveloped by, included in and interacting with the immersive environment \cite{Witmer1998}. 

Thanks to advances in Web technology during the last decade, an increasing number of visual data stories on the Web engage readers with rich and dynamic visuals and interactivity \cite{Zhi2019, Mckenna2017}, and many of them effectively evoke \textit{spatial immersion}, a heightened sense of space or even presence. It is worth mentioning that they rarely require the reader to wear specific systems or devices~(e.g., HMD), but employ a wide range of design techniques, such as virtual diegetic space, depth cue, animated transitions, and camera movements \cite{ryan2015narrative, Zhang2017}. Therefore, the goals of our study, which is to identify common design patterns and to investigate their impact on reading experiences, are much closer to the second view. 

Probably the most prominent intersection of immersive technology and data visualization is in the emergence of immersive analytics, which helps analysts explore and find insights from 3D visualizations beyond traditional 2D displays~\cite{Chen2016, Willett2016, klemm_interactive_2014, bezerianos_perception_2012}. However, at the same time, data storytelling has a different purpose, which is to communicate information through visualized data facts~\cite{Lee2015}. Since Segel and Heer~\cite{Segel2010} first formulated the genres and design space of data storytelling, researchers have studied practitioners' work, uncovering common characteristics and design patterns, such as interactivity, rhetoric, and cinematography~\cite{dove2012narrative, hullman2011visualization, Amini2015}. With the advent of data stories with spatially immersive visuals, we believe that there are open-ended research questions that neither traditional data visualization nor immersive analytics research has fully investigated yet. 

In this paper, we seek to answer two research questions:
\textbf{(RQ1)} What are the design patterns that evoke spatial immersion in visual data story?
To answer the first question, we iteratively visited a wide range of sources, including academic papers, news outlets, online awards, and personal blogs (\S\ref{table:sources of data stories}). Each source was critically reviewed based on criteria of visual data story\cite{Lee2015} and spatial immersion\cite{nilsson2016immersion, Zhang2017}. In consequence, we collected a comprehensive yet compact collection of 23 data stories (\S\ref{sec:case-patterns}) that integrate data-driven facts, the author's message, and an interplay between narrative and data visualization evoking the sense of spatial immersion. Finally, from the collection, we identified six design patterns that give an overview of how practitioners promote spatial immersion.

\textbf{(RQ2)} How does spatial immersion affect the reader's experience?
To answer the second question, we first designed four data stories, each presented in three design variants: \static, \animated, and \immersive. Although the text narratives remain the same across all variants, each has unique visual characteristics as follows. 
First, the \static\ variants have flat designs and scroll (or move to the next page) as a whole, and thus do not evoke any feeling of spatial immersion. In contrast, both the \animated\ and the \immersive\ variants have dynamic layouts so that the readers can easily focus on the corresponding visual elements throughout the narratives. 
On the other hand, while the \animated\ variants employ planar designs akin to the \static\ ones, the \immersive\ variants employ the design patterns to enhance readers' perception of spatial depth.
To compare the impact of each variant on the reading experience, we conducted a within-subjects controlled experiment, where online participants were asked to evaluate each variant they read in terms of five cognitive factors: \emph{interest}, \emph{ease of understanding}, \emph{persuasiveness}, \emph{trustworthiness}, and \emph{curiosity}.
The result suggests that the \immersive\ stories were more interesting and persuasive than the \static\ or \animated\ ones.
However, no single condition was significantly more understandable or trustworthy. To gain a deeper understanding of why participants felt that way, we performed iterative open coding. 
Drawn from the findings, we discuss strengths, weaknesses, and other insights of each variant. 
In addition, we propose design implications, focusing on when and how to utilize the design patterns to get an optimal reading experience. 
Finally, we suggest practical features for evoking spatial immersion in existing libraries, including those for HTML5, chart production, and 3D virtual environment.~(\S\ref{sec:discussion}).

The key contributions of this work are as follows:
\begin{itemize}
    \item By sharing the development process of four data stories with three design variants~(i.e., \static, \animated, and \immersive), we demonstrate considerations and challenges for making data stories with spatial immersion.~(\S\ref{sec:stories}). 
    \item With the result of the within-subjects controlled experiment, we suggest that design patterns for spatial immersion can make a data story more \emph{interesting}, \emph{persuasive}, and better evoke the reader's \emph{curiosity} of the topic~(\S\ref{sec:study}).
    \item We provide design implications for developers and designers in creation of data stories with spatial immersion~(\S\ref{sec:discussion}).
\end{itemize}
\section{Related Work}

\subsection{Spatial Immersion and Engagement} \label{section:spatial_immersion}
The definition of immersion has been interchangeably used with other concepts like ``flow''\cite{Csikszentmihalyi2014}, ``cognitive absorbtion''\cite{Agarwal2000}, and ``presence''\cite{mcmahan2013immersion} depending on their domains and media. However, researchers seemed to have a consensus on immersion as a feeling of being surrounded\cite{mcmahan2013immersion, Witmer1998, ryan2015narrative}. 
Oftentimes, engagement is considered a key characteristic of the immersion\cite{Brown2004, coomans1997towards}, including a sense of being transported into the virtual environment\cite{curran2018factors} and a lack of awareness of time and surrounding environment. 
As elaborated in O'Brien and Tom's work\cite{o2010development}, the engagement has user and system attributes such as aesthetics, emotional response, focused attention, challenge, control, or feedback. These attributes also align with the factors hypothesized to contribute to a sense of space presented in Witmer and Singer's work\cite{Witmer1998}.

Likewise, the researchers also distinguished which elements contribute to such individual experiences of focus and engagement. For example, the immersive experience of reading books and watching films mainly comes with enjoyment of the unfolding narrative and emotional attachment to characters\cite{adams2006fundamentals, ryan2015narrative}. Immersion evoked during game play is more related to active exploration of diegetic world and encountering challenges\cite{ryan2015narrative, mcmahan2013immersion, ermi2007analyzing}. Sometimes, the immersion is just given with rich, enveloping visual and auditory perceptions\cite{Witmer1998}. Among the various elements contributing to immersion, this paper focuses on the deliberate manipulation of the camera and other scene components\cite{Zhang2017}, which facilitate the exploration of virtual spaces with data visualization, creating a pleasurable experience.

\subsection{Design Patterns of Data Stories}\label{section:design_patterns_of_nv}
Data-driven storytelling aims to convey the author's messages by meaningfully connecting findings and visualizations derived from the data\cite{Lee2015}. 
While most research in data visualization focuses on visual analytics~(i.e., discovering insights and data-driven facts), there is limited aid for using data stories to communicate data-driven insights to others\cite{Segel2010}.
As pioneers among researchers investigating data stories, Segel and Heer \cite{Segel2010} analyzed 58 visualizations from journalism, business, and research fields, and formulated a comprehensive design space that includes three ways for categorizing data stories: (1) genres; (2) visual narrative tactics for directing attention, guiding transitions, and orienting audiences; (3) narrative structure tactics such as ordering, interactivity, and messaging.

The narrative aspect plays a critical role in design space. It was examined in detail by Hullman and Diakopoulos\cite{hullman2011visualization}, who expanded the design space with ``visualization rhetoric'' to understand how design techniques impact end-user interpretation, and by Bach et al.\cite{bach2018narrative}, who presented and demonstrated how narrative design patterns can be incorporated into a single data story. 
However, this paper focuses on the visual design elements.
Our work shares a similar scope with Conlen et al.\cite{Conlen2023} in examining data visualization design techniques aimed to engage viewers. 
A key distinction is that their study included data videos and conference talks, while we concentrated on web visual data stories.
Additionally, as noted in Section \ref{section:design_patterns_of_nv}, we focused on mise-en-scène but did not examine editing or sound techniques.
Also, our work aligns with Shi et al.\cite{shi2021communicating} and Lan et al.\cite{lan2021kineticharts} in their focus on animation patterns in data visualization that are designed to achieve impacts on communication and evoking emotions.
However, while their work focused on generalizable animation schemes, we aimed to identify more comprehensive and conceptual strategies which consider style, animation, and composition at the same time. These strategies would serve as design guidance similar to the investigated opening styles and guidelines for implementation by Xu et al.\cite{xu2022wow}.

The identification of design patterns could be followed by both presenting high-level implications for creators\cite{Conlen2023} and a detailed analysis based on factors such as topic or the prevalence of certain patterns over time\cite{mayer2023characterization}. 
However, the primary goal of this paper is to find patterns useful for creating experimental conditions for our controlled experiment rather than conducting a systematic survey, extension of previously presented design spaces, or statistical analysis.

\subsection{Empirical Studies on the Visual Data Stories}\label{section:empirical_studies}
As Tufte\cite{tufte_1986} emphasized the importance of ``ease of understanding'' and ``transparent and objective communication of factual information'' in data visualization, the efficiency of data analytic tasks has long been a primary criterion for evaluating the data visualization. 
On the other hand, data stories are often designed to aid in the memorability of the presented data and message, even if their accuracy and transparency would be lowered as a consequence \cite{Bateman2010}. 
According to Boy et al. \cite{Boy2015_does_it_engage}, ``engagement'' refers to the degree of the user's willingness to invest time and effort to explore further and gain more information, which could be an important emotion for data stories which aim to enhance communication. 
In addition, ``immersion'' has been discussed within the emerging field of immersive visual data stories \cite{isenberg_immersive_2018}.   

In this regard, some researchers conducted empirical studies evaluating the impacts of such engaging and immersive designs on the reader's experience. 
For instance, Bateman et al. \cite{Bateman2010} compared plain charts with embellished charts, and found out that readers felt not only interested in reading them, but also preferred the embellished ones as they were faster and easier to remember.
Barral et al. \cite{Barral2020} tracked the reader's eye behavior to see the impact of an adaptive guide for magazine style data stories.
Also, Romat et al.\cite{romat2020dear} comprehensively assessed impacts of VR experience and personalization on enjoyment, encompassing objective measures on memorability, analyzing verbal expressions, and collecting self-reporting responses using 7-point Likert questionnaires about engagement and VR experience.

It is noteworthy that the online micro-task markets~(e.g., Amazon Mechanical Turk, Prolific) became viable options for such user studies\cite{crowdsourcing_user_studies}.
For example, McKenna et al. \cite{Mckenna2017} conducted a crowdsourced study to assess the impact of the factors about visual narrative flow on reader engagement.
Zhi et al. \cite{Zhi2019} also employed crowdsourcing during their investigation on the impact of layout and interactive linking between text and visualization on the reader's experience in terms of the level of engagement, comprehension, and recall. 
We also conducted a crowdsourced study due to its similarity to real-world scenarios where visual data stories are published to anonymous readers on the web, as well as its ability to collect data at scale.

\subsection{Web-based Data Stories and Scrollytelling}
Data stories employ a wide range of design patterns depending on their genres, types of audience, channels of deployment, storytelling structures, data, and purposes of storytelling \cite{Segel2010}. 
However, this paper pays close attention to web-based data journalism articles for their growing popularity and fast-adoption of the latest web technologies, such as responsive layouts\cite{victor2011explorable}, advanced 3D rendering using WebGL \cite{danchilla2012three}, and high-level languages for authoring data visualizations\cite{bostock2011d3, satyanarayan2016vega}. Enabled by modern web technologies, a few leading groups of online journalists~(e.g., New York Times Data Visualization Lab\footnote{https://open.nytimes.com/the-new-york-times-data-visualization-lab-b52847cf572b}, The Pudding\footnote{https://pudding.cool/}, Wall Street Journal's Graphics Team\footnote{https://graphics.wsj.com/}) invented novel design patterns and successfully integrated them into their narratives\cite{Segel2010}. 
For example, scrollytelling \cite{scrollytelling_2018} animates visual and textual content smoothly as the reader scrolls through an article.
As it has become an integral part of online visual storytelling over a decade\cite{wolf_longforms_journalism_2016}, researchers examined the variety of scrollytelling techniques in terms of visual elements\cite{Seyser_2018}, transition and layout\cite{sultanum2021leveraging}, and what readers control by scrolling\cite{vallandingham_so_nodate}. 
Recently, scrollytelling has been emerging as a standard across both mobile and desktop environments. Alongside various analyses of scrollytelling techniques, research has also been conducted on automating the entire process from data facts extraction to data story creation\cite{lu2021automatic}, as well as proposing authoring tools that support specialized data visualizations\cite{morth2022scrollyvis}.
While scrollytelling is known to be engaging, interactive, and effective in presentation\cite{pettersenscrollytelling}, experts warned about a few drawbacks, including ``scrolljacking''\cite{bostock2011d3}, a lack of clear affordances, and a non-intuitive connection between the user's action and animation\cite{kosara_scrollytelling_2016}.  
\section{Immersive Data Stories on the Web}
\newcolumntype{P}[1]{>{\centering\arraybackslash}p{#1}}
\begin{table*}[t]
\caption{Sources of visual data stories}
\centering{
\begin{tabular}{|l|c|l|}
\hline
Source Name & Source Type & Querying Method\\ \hline

\href{https://www.nytimes.com}{New York Times} & Article Publisher & Feature articles (The Year in Visual Stories and Graphics)\\ \hdashline[0.5pt/5pt]

\href{https://www.theguardian.com/interactive}{The Guardian} & Article Publisher & Articles archived in Interactive Section\\ \hdashline[0.5pt/5pt]

\href{https://graphics.wsj.com/}{Wall Street Journal} & Article Publisher & Projects in Wall Street Journal Graphics\\ \hdashline[0.5pt/5pt]

\href{https://pudding.cool/}{The Pudding} & Article Publisher & Entire works\\ \hdashline[0.5pt/5pt]

\href{https://trends.google.com/trends/story/US_cu_6fXtAFIBAABWdM_en}{Google Trends} & Blog & Search with keyword, "Visualizing Google Data"\\ \hdashline[0.5pt/5pt]

\href{https://flowingdata.com/}{Flowing Data} & Blog & Entire works\\ \hdashline[0.5pt/5pt]

\href{https://www.informationisbeautifulawards.com/}{Information is beautiful Awards} & Award & Work of Winners in all years\\ \hdashline[0.5pt/5pt]

\href{https://www.behance.net/}{Behance} & Portfolio Archive & Search with keyword, "data visualization"\\ \hdashline[0.5pt/5pt]

\href{https://www.verysmallarray.com/}{Very Small Array} & Individual Website & Entire works\\ \hdashline[0.5pt/5pt]

\href{https://www.visualcinnamon.com/}{Visual Ciannamon} & Individual Website & Entire works\\ \hdashline[0.5pt/5pt]

\href{https://truth-and-beauty.net/}{Truth and Beauty} & Individual Website & Entire works\\ \hline

\end{tabular} %
}
\label{table:sources of data stories}
\end{table*}
In this section, we present a literature review on web articles using data visualization that evokes a sense of space. Following Segel and Heer's approach \cite{Segel2010}, the review focuses on two aspects: (1) which visual element increases the sense of spatial immersion, and (2) how stories associate visualization and narrative for better communication. We iteratively visited a growing set of sources on the Web, as listed in Table \ref{table:sources of data stories}, collected target web data stories, and carefully reviewed them to understand how web data stories evoke the feeling of being immersed in different reality with their visuals and narrative. When the iterative collection process did not provide any new insights, we performed an open-coding analysis to identify six common design patterns~(\S\ref{sec:six-design-patterns}). Please note that most of the \textit{design patterns} are neither novel, comprehensive, nor mutually exclusive. Nevertheless, they were helpful for us to describe state-of-the-art data stories on the Web and how they evoke the sense of immersion.

\subsection{Collecting Data Stories that Evoke Spatial Immersion}
\label{sec:case-patterns}
Given that there are no large standard repositories or search engines for web data stories, we employed the network-based snowball sampling method \cite{parker2019snowball} to collect target stories. First, we began our sampling process from the New York Times and Flowing Data, which are two frequently cited resources for web data stories \cite{Segel2010, shu2020makes, ge2020canis}. While collecting data stories from the initial sources, we subsequently added other sources, as summarized in Table~\ref{table:sources of data stories}. In the end, we examined three major media outlets, one interactive data story publisher, two blogs, one award, a portfolio archive service, and three individual websites. 

As the sources comprised a wide range of articles, we defined three criteria to select visual data stories, based on the definition of Lee et al\cite{Lee2015}. First, we filtered out articles that did not contain data-supported facts. Second, we collected articles containing at least one intended message. 
Lastly, even if an article contained data facts and an intended message, we dropped it unless it had an interplay between narrative and visualization. 
Using these criteria, we excluded simple chart images without written explanations or annotations, articles that discuss data-driven content without including charts, and cases where charts and data-driven facts are presented but viewers must rely on free exploration to gather information without any narrative guidance.

Subsequently, three researchers evaluated each visual data story in terms of whether they evoke the feeling of spatial immersion, using the following criteria based on previous research~\cite{nilsson2016immersion, Zhang2017}. First, the viewpoint or elements of the story must move continuously and dynamically to engage the readers. With this criterion, we excluded data stories that contain static 3D graphics but did not effectively engage us (as readers). Second, the layout of the story must have a spatial composition that helps the readers feel as if they walk into another 3D environment and interact with the surroundings. Thus, we excluded articles that evoked non-spatial immersion. Lastly, the visual elements of each story must facilitate the reading experience and understanding of the story so that readers can easily follow the narrative.

In order to prevent redundant cases in our collection, whenever a new visual data story was found, we checked if there was a story in the collection that used the same techniques on a similar group of spatial compositions~(e.g., the transformation of a heatmap into a 3D representation using bars rising above a world map). Finally, a collection of 23 data stories was made, as listed in Table~\ref{table:visual data story cases}.

\subsection{Extracting Common Design Patterns}
\label{sec:design-patterns}
\newcolumntype{P}[1]{>{\centering\arraybackslash}p{#1}}
\begin{table*}[t]
\caption{Collected visual data stories with design patterns}
\setlength{\tabcolsep}{1pt}
\renewcommand{\arraystretch}{1.2}
\centering{
\begin{tabular}{|m{10.0cm}|P{1.0cm};{0.5pt/5pt}P{1.0cm};{0.5pt/5pt}P{1.0cm};{0.5pt/5pt}P{1.0cm};{0.5pt/5pt}P{1.0cm};{0.5pt/5pt}P{1.8cm}|}
\multicolumn{1}{c|}{} & \multicolumn{6}{c|}{\textbf{Design Pattern}} \\ 
\multicolumn{1}{c|}{\textbf{}} & \multicolumn{1}{c}{\textbf{1}} & \multicolumn{1}{c}{\textbf{2}} & \multicolumn{1}{c}{\textbf{3}} & \multicolumn{1}{c}{\textbf{4}} & \multicolumn{1}{c}{\textbf{5}} & \multicolumn{1}{c|}{\textbf{6}} \\ 
\multicolumn{1}{c|}{Source} & 
\multicolumn{1}{c}{\includegraphics[width=0.06\textwidth]{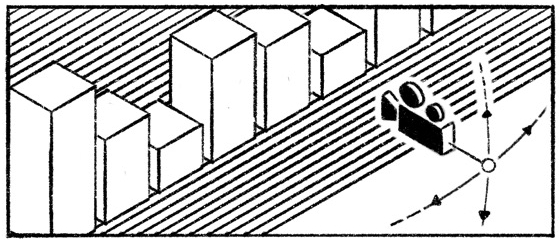}} & 
\multicolumn{1}{c}{\includegraphics[width=0.06\textwidth]{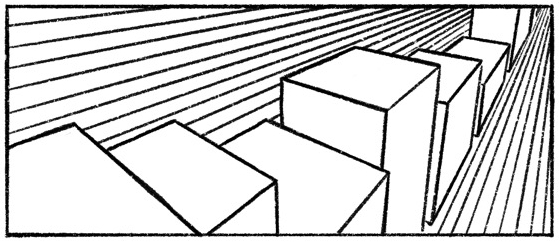}} & 
\multicolumn{1}{c}{\includegraphics[width=0.06\textwidth]{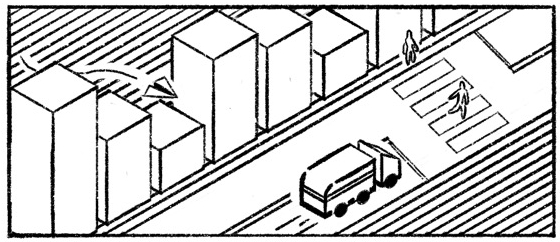}} & 
\multicolumn{1}{c}{\includegraphics[width=0.06\textwidth]{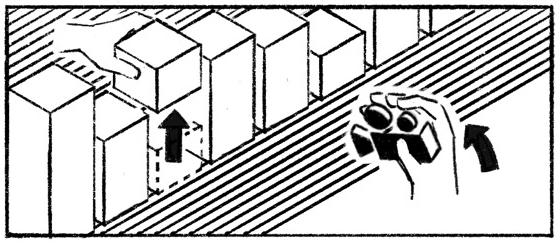}} & 
\multicolumn{1}{c}{\includegraphics[width=0.06\textwidth]{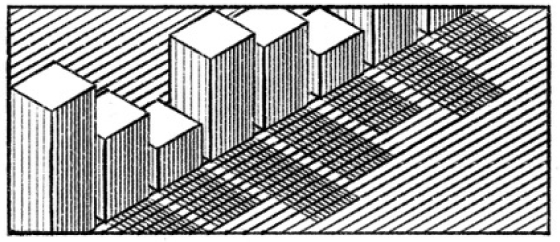}} & 
\multicolumn{1}{c|}{\includegraphics[width=0.11\textwidth]{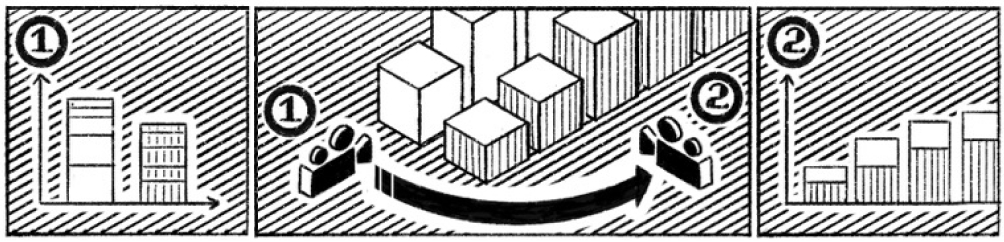}}
\\ 
\cline{1-7}

\href{https://www.nytimes.com/interactive/2018/05/09/nyregion/subway-crisis-mta-decisions-signals-rules.html}{[S1]} How 2 M.T.A. Decisions Pushed the Subway Into Crisis & \ding{52} & \ding{52} & \ding{52} & \ding{52} & ~ & ~ \\ \hdashline[0.5pt/5pt]

\href{https://www.nytimes.com/interactive/2021/09/15/nyregion/empire-state-building-reopening-new-york.html}{[S2]} Why the Empire State Building, and New York, May Never Be the Same & \ding{52} & \ding{52} & \ding{52} & ~ & \ding{52} & \ding{52} \\ \hdashline[0.5pt/5pt]

\href{https://buildinghop.es/}{[S3]} Google Building Hopes & \ding{52} & ~ & \ding{52} & \ding{52} & \ding{52} & ~ \\ \hdashline[0.5pt/5pt]

\href{https://pudding.cool/2017/10/satellites/}{[S4]} Seeing Earth from Outer Space & \ding{52} & ~ & \ding{52} & ~ & \ding{52} & ~ \\ \hdashline[0.5pt/5pt]

\href{https://pudding.cool/2018/12/3d-cities-story/}{[S5]} Population Mountains & \ding{52} & \ding{52} & ~ & ~ & ~ & \ding{52} \\ \hdashline[0.5pt/5pt]

\href{https://www.nytimes.com/interactive/2015/03/19/upshot/3d-yield-curve-economic-growth.html}{[S6]} A 3-D View of a Chart That Predicts The Economic Future: The Yield Curve & \ding{52} & \ding{52} & ~ & \ding{52} & ~ & \ding{52} \\ \hdashline[0.5pt/5pt]

\href{https://www.nytimes.com/interactive/2021/01/28/opinion/climate-change-risks-by-country.html}{[S7]} Every Country Has Its Own Climate Risks. What’s Yours? & \ding{52} & \ding{52} & ~ & ~ & ~ & \ding{52} \\ \hdashline[0.5pt/5pt]

\href{https://pudding.cool/2018/10/city_3d/}{[S8]} Human Terrain & \ding{52} & \ding{52} & ~ & \ding{52} & ~ & ~ \\ \hdashline[0.5pt/5pt]

\href{https://pudding.cool/2018/03/neighborhoods/}{[S9]} A Tale of Two Cities & \ding{52} & ~ & ~ & \ding{52} & ~ & ~ \\ \hdashline[0.5pt/5pt]

\href{https://www.nytimes.com/interactive/2021/01/14/climate/hottest-year-2020-global-map.html}{[S10]} Where 2020's Record Heat Was Felt the Most & \ding{52} & ~ & ~ & ~ & ~ & ~ \\ \hdashline[0.5pt/5pt]

\href{https://www.nytimes.com/interactive/2021/05/24/us/tulsa-race-massacre.html}{[S11]} What The Tulsa Race Massacre Destroyed & \ding{52} & ~ & ~ & ~ & \ding{52} & \ding{52} \\ \hdashline[0.5pt/5pt]

\href{https://www.nytimes.com/interactive/2021/06/30/opinion/environmental-inequity-trees-critical-infrastructure.html}{[S12]} Since When Have Trees Existed Only for Rich Americans? & \ding{52} & ~ & ~ & ~ & ~ & ~ \\ \hdashline[0.5pt/5pt]
\href{https://pudding.cool/2018/07/women-in-congress/}{[S13]} We mapped out the road to gender parity in the House of Representatives & \ding{52} & ~ & ~ & ~ & ~ & ~ \\ \hdashline[0.5pt/5pt]
\href{https://pudding.cool/2018/02/stand-up/}{[S14]} Structure of Stand Up Comedy & \ding{52} & ~ & ~ & ~ & ~ & ~ \\ \hdashline[0.5pt/5pt]
[S15] Tracing the K-POP WAVE & ~ & \ding{52} & \ding{52} & ~ & ~ & ~ \\ \hdashline[0.5pt/5pt]
\href{https://pudding.cool/2017/01/making-it-big/}{[S16]} The Unlikely Odds of Making It Big & ~ & \ding{52} & \ding{52} & ~ & ~ & ~ \\ \hdashline[0.5pt/5pt]
\href{https://www.nytimes.com/interactive/2021/03/10/opinion/covid-vaccine-lines-states.html}{[S17]} Who’s Next in Your State’s Vaccine Line? & ~ & \ding{52} & ~ & ~ & \ding{52} & ~ \\ \hdashline[0.5pt/5pt]
\href{https://www.nytimes.com/interactive/2021/07/30/sports/olympics/olympic-running.html}{[S18]} How Speed and Distance Dictate How Olympians Run & ~ & \ding{52} & ~ & ~ & ~ & ~ \\ \hdashline[0.5pt/5pt]
\href{https://www.theatlantic.com/theplatinumpatients/}{[S19]} The Platinum Patients & ~ & ~ & \ding{52} & ~ & ~ & ~ \\ \hdashline[0.5pt/5pt]
\href{https://pudding.cool/2017/04/beer/}{[S20]} Craft beer — so hot right now. But what city is the microbrew capital of the US? & ~ & ~ & ~ & \ding{52} & ~ & \ding{52} \\ \hdashline[0.5pt/5pt]
\href{https://pudding.cool/2021/03/love-and-ai/}{[S21]} Nothing Breaks Like A.I. Heart & ~ & ~ & ~ & \ding{52} & ~ & ~ \\ \hdashline[0.5pt/5pt]
\href{https://pudding.cool/2017/03/punk/}{[S22]} Crowdsourcing the Definition of "Punk" & ~ & ~ & ~ & ~ & ~ & \ding{52} \\ \hdashline[0.5pt/5pt]
\href{https://pudding.cool/2018/08/retraining/}{[S23]} Why the tech sector may not solve America’s looming automation crisis & ~ & ~ & ~ & ~ & ~ & \ding{52} \\
\hline
\end{tabular} %
}
\label{table:visual data story cases}
\end{table*}
Each story in the collection has its own way of utilizing the interplay of visual and narrative elements. To find common patterns, we analyzed them based on the design space created by Segel and Heer\cite{Segel2010}, which encompasses narrative tactics with/without visualization. 
In addition, we explore the perceptual and structural characteristics of the visual composition as well.
To sum up, our analysis aims to examine (1) how visual elements such as configuration and layout of the stories and characteristics of charts contribute to the reader's feeling of spatial immersion and (2) how narrative elements (e.g., interactivity, transition, navigation, annotation) facilitate communication.

During the examination of these elements and the counting of their frequencies, we discovered some valuable findings that characterize the designs of the stories. 
First, even though most of the stories employ flat 2D charts and abstract visual encoding, they effectively evoke spatial immersion when animated in virtual 3D environments. For example, some of the virtual environments depict a wide range of scenes existing in the real world, ranging from city blocks~(S11) to the Empire State building~(S2), encompassing aerial perspectives of particular areas~(S7, S8, S12), outer space~(S4), and the globe~(S10). The other virtual environments consist of rather abstract visuals such as chart axes and markers~(S6, S13, S16, S20, S22, S23), diagrams and text blocks~(S14, S21), and orbits with rotating objects~(S3, S16).

The collected stories give readers limited freedom to modify the visual encoding of existing charts or create their own charts from scratch. Instead, they tend to change content and representations along the linear narrative predefined by the authors. 
For example, in S20, as readers scroll to the text boxes discussing New York, Santa Rosa, and the entire USA, the story triggers predetermined camera panning and zoom effects to highlight the corresponding region on the map.
As another example of linear narrative, S2 evenly divides entire scroll progress into tiny steps and maps all frames of predefined animation one-by-one to these steps.

Lastly, most of the stories in our collection align the text blocks and the corresponding visualizations to provide meaningful connections. For example, S18 frequently uses text labels and annotations for detailed information and shows various meaningful layouts of text and visual components. As another example, many stories begin with a lengthy text, but as readers scroll down the page, videos, charts, and other visual elements appear and interleave with small text boxes.
In many stories, including S18, these text boxes are visually distinguished from the other visual elements using shadows, background colors, and borders.

\subsection{Six Design Patterns for Promoting Spatial Immersion}
\label{sec:six-design-patterns}
Based on the above findings, this section presents six \textit{design patterns}~(i.e., \texttt{DP}) which further elaborate on the anticipated sense of space and how it was promoted in the examples provided. 
Please note that the \texttt{DPs} are neither exhaustive nor mutually exclusive by our intention. In fact, each story has at least one and up to five \texttt{DPs} as presented in Table \ref{table:visual data story cases}. Some \texttt{DPs} overlap each other and there may be patterns that we missed. Nevertheless, \texttt{DPs} are more useful to get an overview of how practitioners promote spatial immersion and to create the \immersive\ conditions in \S\ref{sec:stories}. Later in this section, we elaborate on the role, limitations, and use cases of each \texttt{DP}. 

\begin{wrapfigure}{l}{0.18\textwidth}
\centering
\includegraphics[width=0.18\textwidth]{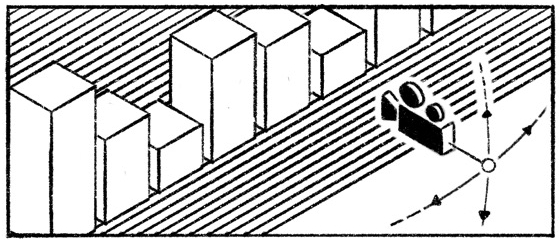}
\vspace{-22pt}
\end{wrapfigure}
\paragraphHeading{DP1: \dpA} helped us to feel spatially immersed in the scene. 14 out of 23 data stories in our collection employ a variety of camera shots~(e.g., wide, close-up, point-of-view shots) and transitions~(e.g., tracking, hovering, zooming in and out), which seem to be heavily influenced by the film industry. For example, in S2, the path of the virtual camera makes readers feel as if they are hovering around the Empire State Building. Notably, only five stories use 3D charts, while the other nine stories use 2D charts placed in virtual 3D spaces - but they all successfully create the feeling of spatial immersion. For example, S13 contains 2D charts, but as readers scroll down, the virtual camera would zoom and pan to the area of narrative focus. This example involves a popular design technique on the web, called scrollytelling \cite{scrollytelling_2018}. However, the main point of \texttt{DP1} is not the scroll-driven narrative control but the creative use of a virtual camera that flies along the author-directed narrative path in 3D space - which helps readers accurately interpret the meaning of individual 2D charts, and get immersed into the narrative space without friction.

\begin{wrapfigure}{l}{0.18\textwidth}
\centering
\vspace{-12pt}
\includegraphics[width=0.18\textwidth]{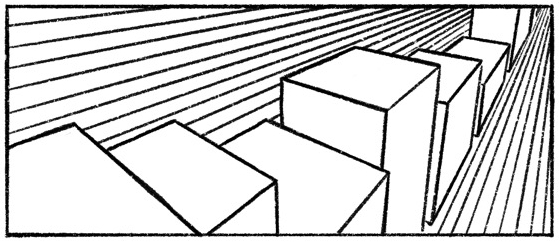}
\vspace{-22pt}
\end{wrapfigure}
\paragraphHeading{DP2: \dpB} is a design pattern of encoding colors, shapes, animations, and interactivity of a chart's elements in analogy to a specific phenomenon or space in the narrative. To give a few examples, S5, S7 and S8 encode the population density of each geographical location as the elevation of mountain range in 3D. S1 features a conventional line chart that represents train schedules, but at the same time the lines depict railroad tracks for animated train icons. In S18, normal forces generated by an athlete’s steps are visualized as downward arrows that grow from the ground~(in accordance with the video clip of the athlete). The main purpose of such encoding is to reduce the reader's cognitive workload by mapping real world objects~(e.g., geographical location, railroad track, athlete's step) to abstract graphical elements~(e.g., chart bars and lines). However, as a side effect of the intuitive mapping, readers would have the sense of being transported while viewing charts. 

\begin{wrapfigure}{l}{0.18\textwidth}
\centering
\vspace{-12pt}
\includegraphics[width=0.18\textwidth]{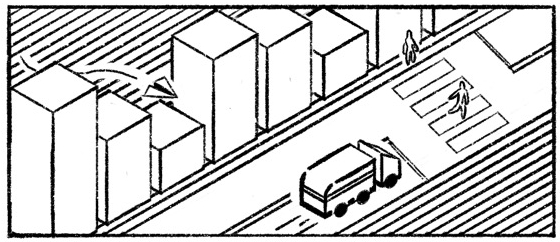}
\vspace{-22pt}
\end{wrapfigure}
\paragraphHeading{DP3: \dpC} usually do not convey meaningful information alone, but aided grabbing attention and was indirectly helpful for feeling the virtual space. For example, workers moving on the floor plan~(S2) and trains running and stopping on railroad tracks~(S1) depict the objects' natural movement, and thus make readers feel immersed in the virtual spaces. Additionally, natural movements can simulate gravity, friction, and other rules of physics. For instance, S18 depicts music bands as dots that smoothly rotate along the concentric circles, making readers indirectly feel the gravity of the planet. Such feelings not only make a virtual space feel more realistic but also help readers get immersed in the space. 

\begin{wrapfigure}{l}{0.18\textwidth}
\centering
\includegraphics[width=0.18\textwidth]{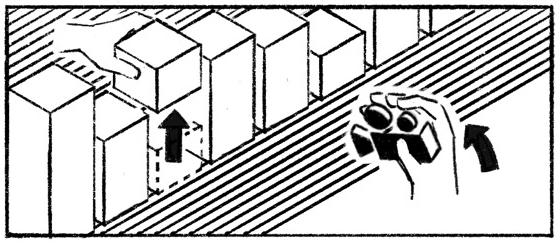}
\vspace{-22pt}
\end{wrapfigure}
\paragraphHeading{DP4: \dpD} enables free exploration of the virtual space, and thus directly promotes the sense of spatial immersion. While the pattern is extremely common in the domain of games, VR content, and other immersive media; the majority of web-based visual data stories use cinematic techniques~(\texttt{DP1}) more frequently, in order to guide readers along the predefined narratives. Indeed, only 6 out of 23 stories in our collection support direct manipulations for zooming and panning cameras~(S6 and S9), and modifying visual components via interactive UI components such as slider, button, text input, interactive heatmap, and drop-down menu~(S1, S3, S20, and S21). 

\begin{wrapfigure}{l}{0.18\textwidth}
\centering
\vspace{-12pt}
\includegraphics[width=0.18\textwidth]{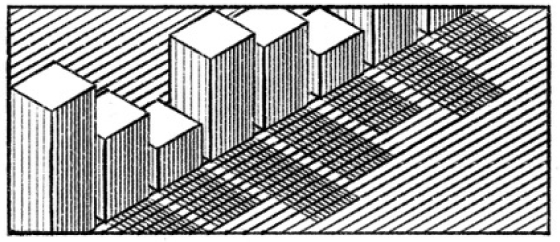}
\vspace{-22pt}
\end{wrapfigure}
\paragraphHeading{DP5: \dpE} is achieved with various rendering techniques, such as image and texture maps, lighting, and shadow effects. While abstract and geometric shapes in solid colors have been a standard look-and-feel of data visualizations \cite{wainer_1984, kosslyn_1985, tufte_1986, Zacks1998ReadingBG}, five out of 23 data stories use realistic charts and elements. For example, three stories applied texture maps on simplified 3D models of the Empire State Building (S2), the Earth from outer space (S4), and a street view (S11). Another case (S3) leverages directional light in a 3D environment. Lastly, S17 employs realistic line drawings of people standing in line, instead of simple bar charts or icons. Notably, \texttt{DP1} and this pattern are used frequently (4 out of 5 times) in conjunction, as they both make virtual spaces more realistic. Like in \texttt{DP1} and \texttt{DP3}, realistic appearance can make readers feel comfortable by leveraging their prior knowledge of physical objects and spaces. 

\begin{wrapfigure}{l}{0.28\textwidth}
\centering
\vspace{-12pt}
\includegraphics[width=0.28\textwidth]{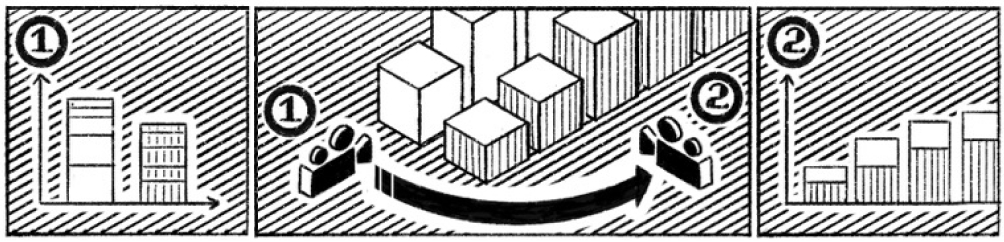}
\vspace{-22pt}
\end{wrapfigure}
\paragraphHeading{DP6: \dpF} is a transition technique to switch between 2D~(planar) and 3D~(perspective) views in a single virtual space. A typical use case is (1) to show a 2D view at key frames, and (2) to rotate the chart or move the camera in 3D while transitioning to the next key frame. For instance, while S6 contains a single 3D patch, for a few key frames the camera smoothly moves to the exact positions~(e.g., top, front, side view) where the 3D patch is rendered as corresponding 2D charts. Similarly, S20 and S23 switch between 1D and 2D views by adding and removing  additional axes. In contrast to conventional data visualizations, dynamic dimensionality can engage readers by guiding them through the narrative, as well as reduce their cognitive load for matching elements in multiple charts.
\section{Data Story Design}
\label{sec:stories}
In this section we present the development of four data stories~(DS), each featuring three design variants: \static, \animated, and \immersive. We set the three design variants to control the effect of the design patterns on the reading experience. The \static\ stories feature charts and text components, which are clearly separated in 2D planes and scrolled cohesively as the reader navigates the story. However, in the \animated\ and the \immersive\ stories, charts and text move in any direction and/or change properties. For example, they are often positioned in a meaningful way so that readers can quickly find the corresponding charts and text. Also, changes of visual attributes such as zoom level, color, and opacity draw the reader's attention to the specific part of the visualization. In addition, while the \animated\ stories do not employ any of the design patterns, the \immersive\ data stories leverage different groups of \texttt{DPs} based on the specific feeling of spatial immersion they aim to achieve.

We created the four data stories with the following considerations in mind. First, in order to obtain generalizable and unbiased findings, the stories encompass distinct subject domains, narrative structures, and combinations of the design patterns. To minimize potential biases from each participant's perspective and interest, the stories convey objective information in the domains of economy, history, and health care. 

Second, each story consists of a single scene, which is shorter and simpler than typical web articles but allows participants to read quickly and answer questions. A scene features a combination of two to four design patterns. Hence, every pattern is used at least once, yet no story was overcrowded with visual effects. For example, DS1 simply uses 3D charts and camera movements~(\texttt{DP1}) for highlighting and dimension conversion~(\texttt{DP6}). DS2 visualizes a bar chart as an iceberg slowly swinging on the ocean~(\texttt{DP2}, \texttt{DP3}) with a realistic texture~(\texttt{DP5}). DS3 features a first person perspective camera~(\texttt{DP1}) moving along the roller-coaster-like presentation of stock market trends~(\texttt{DP2}), where the camera frequently switches between planar and perspective views~(\texttt{DP6}). Lastly, DS4 features camera flying~(\texttt{DP1}) through the valley of heatmap~(\texttt{DP2}), rendered with lighting and shadow effects~(\texttt{DP5}). It also provides free exploration at the end of the story~(\texttt{DP4}).


%



Lastly, for a fair comparison between the design variants, we developed the stories using the same JavaScript library, Three.js. All the stories are optimized to run in any Webkit-based desktop browser\footnote{Chrome, Safari, and Firefox}. We highly recommend trying the three versions on our website \footnote{https://immerwebds.github.io/}.
 


\subsection{DS1: Does Urbanization Affect Food Consumption?}
\label{sec:stories:DS1}
\begin{figure*}[tbp!]
    \centering
    \includegraphics[width=0.66\textwidth,keepaspectratio]{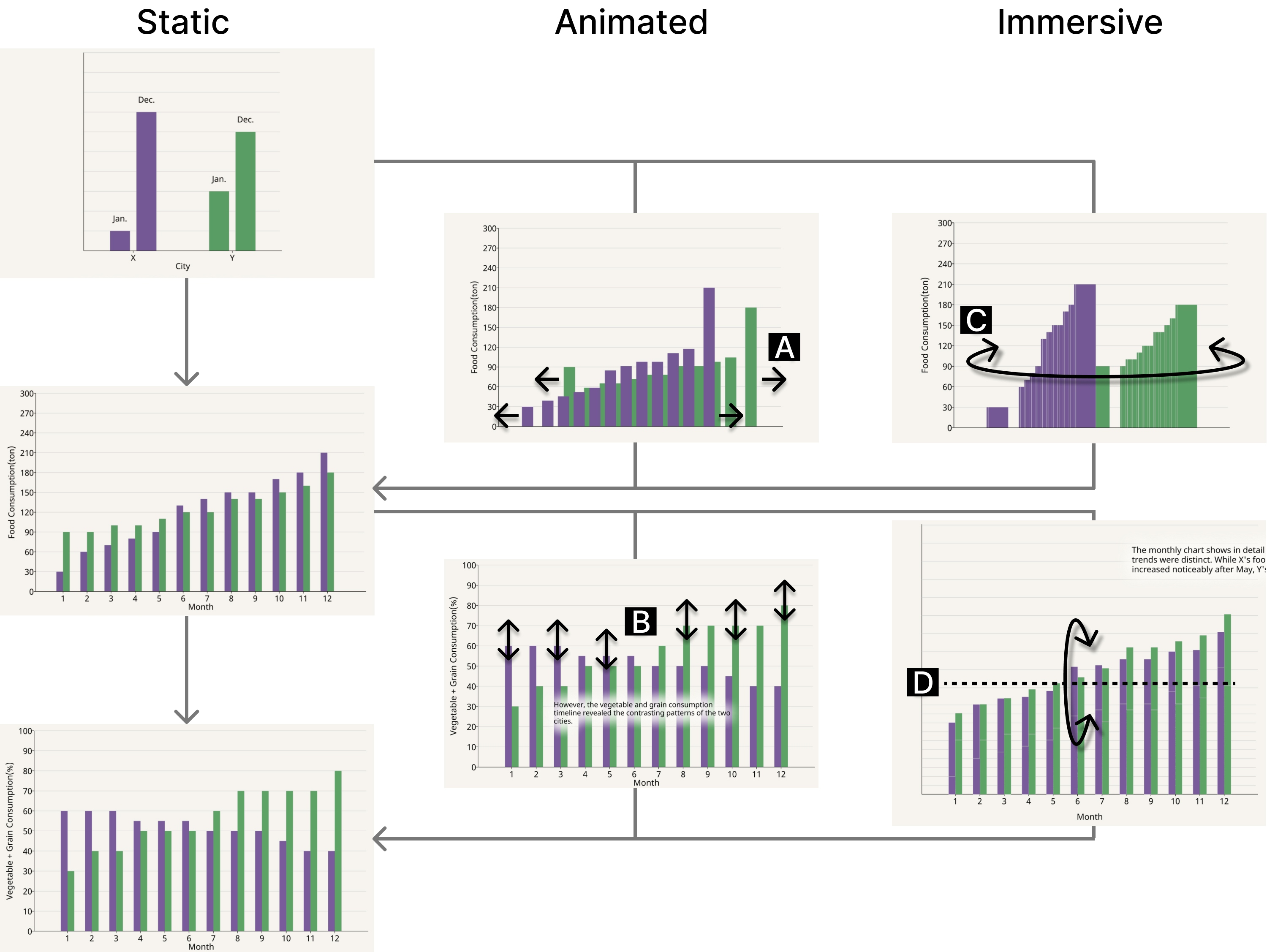}
    \caption{DS1 shows bar charts comparing food consumption of two cities. To visually connect scenes in the narrative, the \emph{animated}~(A, B) continuously transform and scale the bars in 2D, while the \emph{immersive}~(C, D) rotates and moves the camera in 3D.}
    \vspace{-8pt}
    \label{fig:DS1}
\end{figure*}
DS1 features bar charts that compare the food consumption of two cities. It begins with a faceted bar chart that compares the two cities' total food consumption and shows a significant increase from January to December.
Second, it uses a grouped bar chart of total consumption throughout the year, highlighting an intersecting point of consumption between the cities in June.
Lastly, the grouped bar chart switches the y-axis encoding to consumption of vegetable and grain, and shows that one city has a decreasing pattern while the other has an increasing trend line. 
In sum, DS1 hypothesizes that urbanization might have an increase in total consumption, but did not seem to affect the consumption of plant-based foods.

The \static\ variant shows three bar charts accompanied by text paragraphs. However, the \animated\ and \immersive\ continuously manipulate a single chart or the camera as the readers scroll down the page. Both variants animate the grouped bars of the twelve months in different ways. 
While changing the faceted chart into the grouped bar chart, the \animated\ story employs flat animation that widens the gap between January and December, and grows the other ten months from the bottom~(Figure \ref{fig:DS1}A).
However, the \immersive\ one rotates the camera~(Figure \ref{fig:DS1}C) to make the readers perceive the following transition that switches the scene into 3D~(\texttt{DP6}).
In the last scene, the stories change the y-axis encoding from total consumption to the plant-based food ratio.
Again, the \animated\ variant uses flat animation that adjusts the heights of the bars to change the y-axis values~(Figure \ref{fig:DS1}B). On the other hand, the \immersive\ rotates the camera from front to top view, maintaining the x-axis and switching the y-axis with the z-axis~(Figure \ref{fig:DS1}D).

\subsection{DS2: Why Happiness is Becoming More Expensive}
\label{sec:stories:DS2}
\begin{figure*}[tbp!]
    \centering
    \includegraphics[width=0.66\textwidth,keepaspectratio]{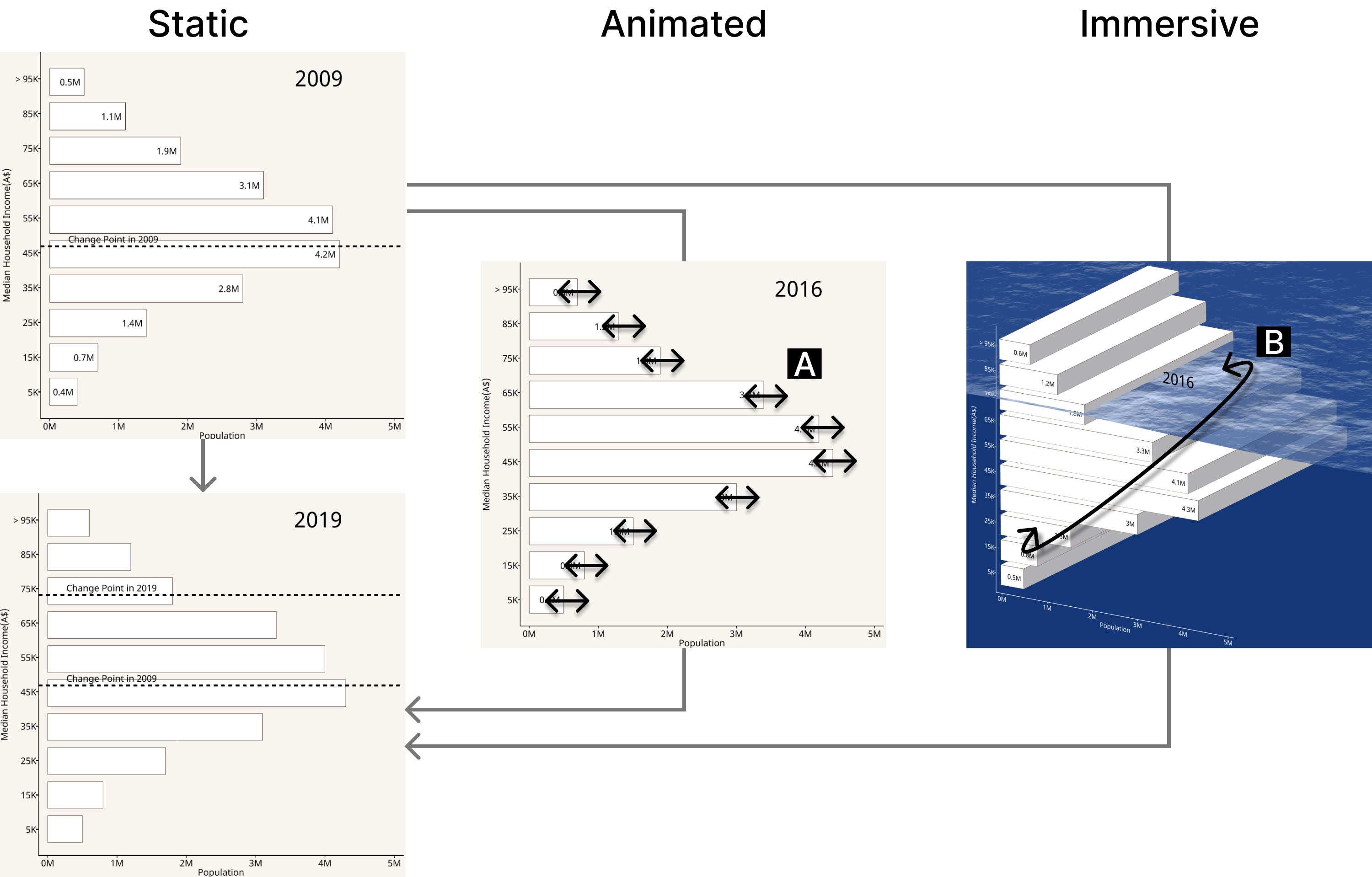}
    \caption{DS2 visualizes how the household income distributions changed between 2009 and 2019, in relation to the change points. While the \emph{animated}~(A) visualizes the change as 2D animations, the \emph{immersive}~(B) leverages the realistic ocean surface in 3D.}
    \vspace{-8pt}
    \label{fig:DS2}
\end{figure*}
The narrative of DS2 is based on an article \cite{happiness_2021}, whose main message is about the stagnated household income and increasing ``change point'', which refers to the minimum amount of income that matters less for happiness.
As illustrated in Figure \ref{fig:DS2}, the story begins with a bar chart visualizing the distributions of household income population in 2009 and the horizontal line that indicates the change point. The following scene shows the distribution in 2019 and two lines indicating the change points of 2009 and 2019. The story concludes by telling that an increasing number of people struggle to disentangle their happiness from monetary concerns.

All three variants compare household income distributions and change points in 2009 and 2019.
While the \static\ story displays the distributions and change point lines of 2009 and 2019 in separate charts, the \animated\ and \immersive\ variants use single bar charts that smoothly interpolate 2009 and 2019 by animating each bar and updating corresponding values~(Figure \ref{fig:DS2}A).
Likewise, the \immersive\ leverages intuitive mapping between the change point and household income to elicit an anxiety~(\texttt{DP2}). In particular, it depicts the ocean surface with blue shimmering animations. The household income floats on the ocean surface, implying that people below the change point are under water~(\texttt{DP5}). These floating data points and the surface of the water move smoothly and calmly as a cold ocean and an iceberg~(\texttt{DP3}). Finally, we exposed the ocean surface through a camera transition from 2D orthographic to 3D perspective view, in order to dramatically catch the readers' attention to the 3D virtual space.

\subsection{DS3: The Stock Market’s COVID Pattern}
\label{sec:stories:DS3}
\begin{figure*}[tbp!]
    \centering
    \vspace{16pt}
    \includegraphics[width=0.66\textwidth,keepaspectratio]{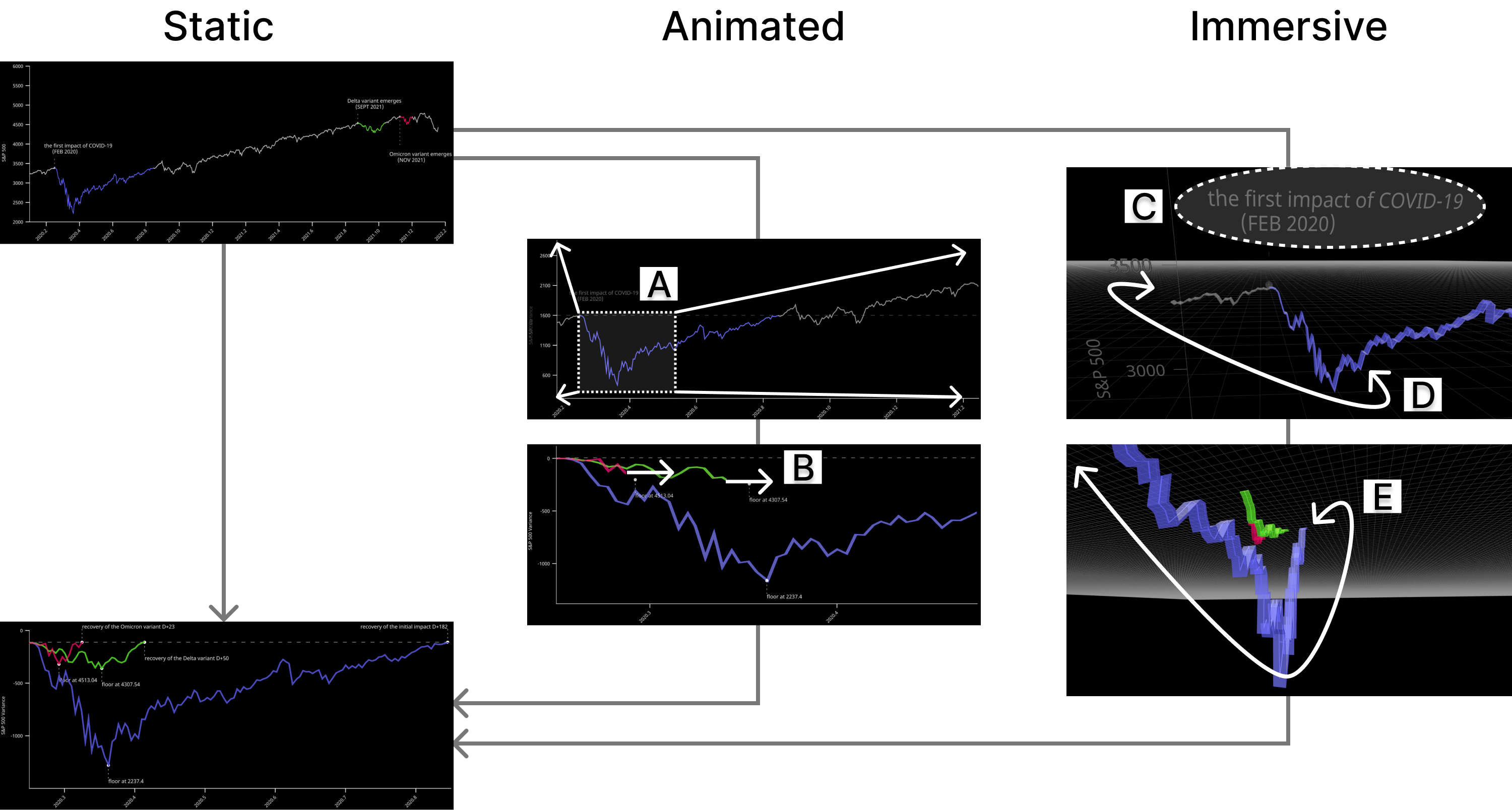}
    \caption{DS3 compares the three economic downturns during the pandemic, and how quickly they got recovered. While the \emph{animated} utilizes (A) 2D data zoom and (B) line animations, the \textit{immersive} uses (C) fading out unimportant elements, (D) a flying camera in 3D space, and (E) dramatic 3D perspectives.}
    \vspace{-8pt}
    \label{fig:DS3}
\end{figure*}
The narrative of DS3 is an excerpt from an article\footnote{https://www.nytimes.com/interactive/2021/12/07/business/omicron-stock-market-covid.html} about major economic downturns during the COVID pandemic. 
As shown in Figure \ref{fig:DS3}, DS3 begins with a line chart, whose x-axis is a time and the y-axis is a stock market index. 
The chart then brings three major downturns to the origin and omits the rest data points, employing the x-axis as the duration and the y-axis as the index variance. The three lines start at the same location so that readers can easily compare the duration and drops. The story concludes with an explanation of the economy gradually gaining resilience against downturns. 

The \static\ variant introduces the topic in the first scene and then compares the three downturns and reports the insights in the following scenes. In contrast, the \animated\ uses a zoom-in effect to continuously show the x-axis change~(Figure \ref{fig:DS3}A), altering the tick positions and values based on the zoom level. In addition, the two lines depicting the corresponding downturns spread out from the same origin point~(Figure \ref{fig:DS3}B), reaching the deepest and farthest points step by step. Influenced by Kenny and Becker's VR content\footnote{http://graphics.wsj.com/3d-nasdaq/}, the \immersive\ uses a roller coaster metaphor to promote a sense of space so that readers can intuitively grasp the story. In other words, the virtual camera dynamically follows the roller coaster-like 3D line~(\texttt{DP1, DP2}), and temporarily stops and looks around, at the lowest and farthest points~(Figure \ref{fig:DS3}E). Meanwhile, the axes and annotations become invisible so that the reader can focus on visual perception. 

\subsection{DS4: Mortality Rates of France in History}
\label{sec:stories:DS4}
\begin{figure*}[htbp!]
    \centering
    \includegraphics[width=0.80\textwidth,keepaspectratio]{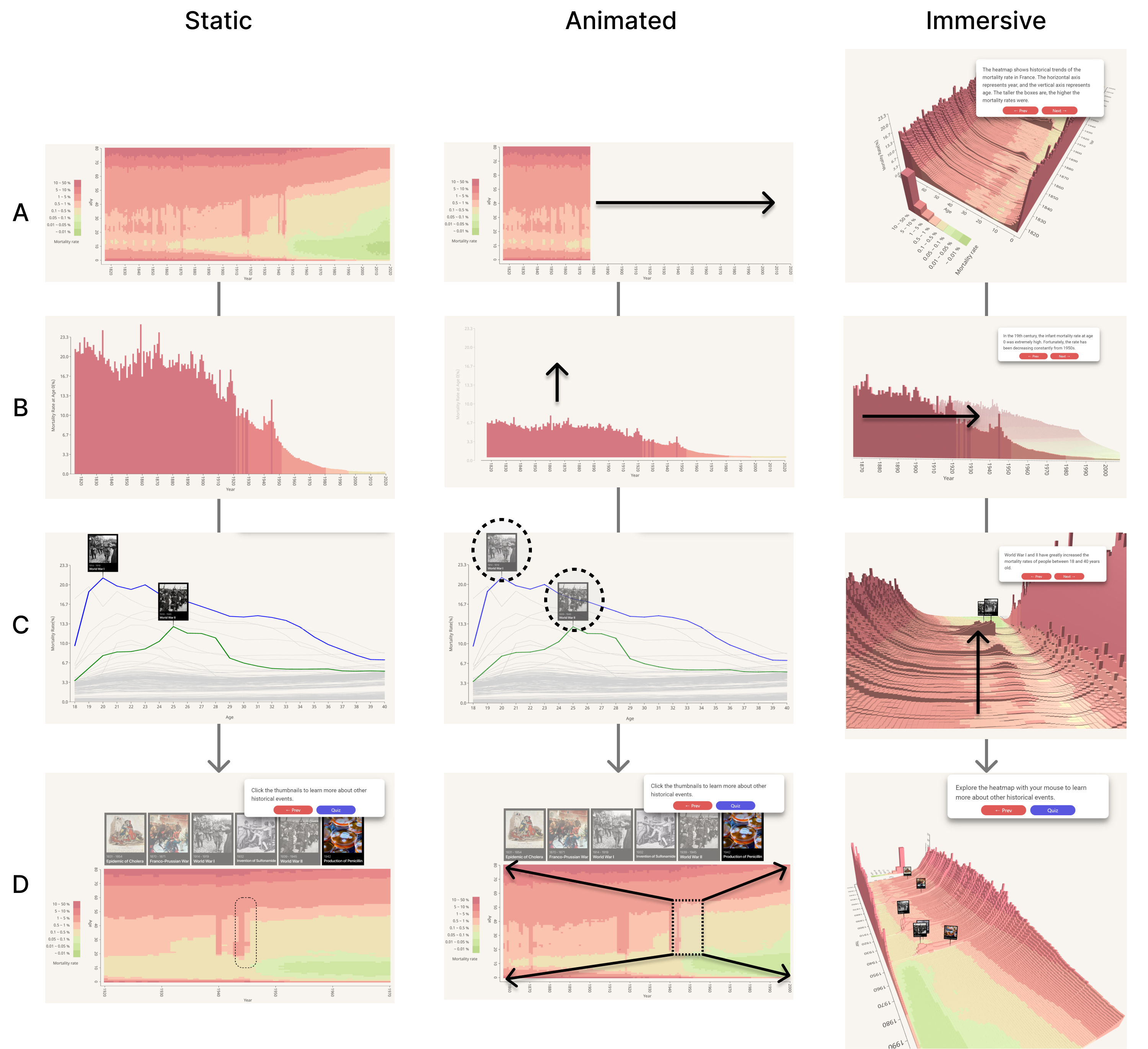}
    \caption{DS4 visualizes the impact of historical events on the mortality rates in France. (A) When introducing the heatmap, the \animated\ variant gradually reveals the 2D heatmap, while the \immersive\ uses the camera flying from the top to the quarter view. (B) To visualize the trend of infant mortality rates, the bars in the \animated\ vertically grow, but the camera in the \immersive\ flies from left to right. (C) To highlight mortality rates for specific years, the \static\ and the \animated\ use the color and thickness encoding, but the \immersive\ one simply moves the camera to the appropriate position. (D) In the \static\ and \animated\ stories, readers can explore the historical events by clicking thumbnails, but the \immersive\ requires readers directly control the position of the camera. 
    }
    \vspace{-8pt}
    \label{fig:DS4}
\end{figure*}
The fourth data story highlights the historical events that affected the mortality rates in France from the 1800s. Such events include the two World Wars and the Cholera epidemic, which dramatically increased mortality rates, and the inventions of sulfonamides and penicillin, which lowered the rates. The story first introduces a heatmap to explain the axes and the color encoding. Then it displays charts, text, and associated views for each historical event group. At the end of the story, the readers explore the historical events with the aid of the annotations.

As shown in Figure \ref{fig:DS4}, both the \static\ and \animated\ stories visualize the mortality rates in a 2D heatmap~(x-axis:year, y-axis:age, color:mortality rate), 2D bar charts, and line charts. The \animated\ one employs a variety of animated transitions: (1) data points are revealed chronologically~(Figure \ref{fig:DS4}A), (2) opacity of texts, axes, and images fade in\&out~(Figure \ref{fig:DS4}B and \ref{fig:DS4}C), (3) gray lines emerge first, then two lines are colored blue and green~(Figure \ref{fig:DS4}C), and (4) camera pans and zooms in\&out~(Figure \ref{fig:DS4}D).
Inspired by S5 and S8, the \immersive\ variant encodes the mortality rates with elevation and colors of the 3D terrain, making the visualization look like a deep valley~(\texttt{DP2}). DS4 heavily employs camera movements compared to other DSs. For example, the camera overlooks the valley at the beginning ~(Figure \ref{fig:DS4}A), follows the trend line of the decreasing mortality~(Figure \ref{fig:DS4}B), and then swiftly files through the valley~(Figure \ref{fig:DS4}C). To make the visualization look realistic~(\texttt{DP5}), we carefully set the lighting, shadow, and camera settings. However, DS4 also allows the reader to manually change camera position and zoom level using a mouse or track pad~(\texttt{DP4}). In the last scene, the thumbnails invite the reader to click on the image to see the details of the key events. 
\section{Crowdsourced User Study}
\label{sec:study}

\begin{figure*}[t]
    \centering
    \includegraphics[width=0.75\textwidth,keepaspectratio]{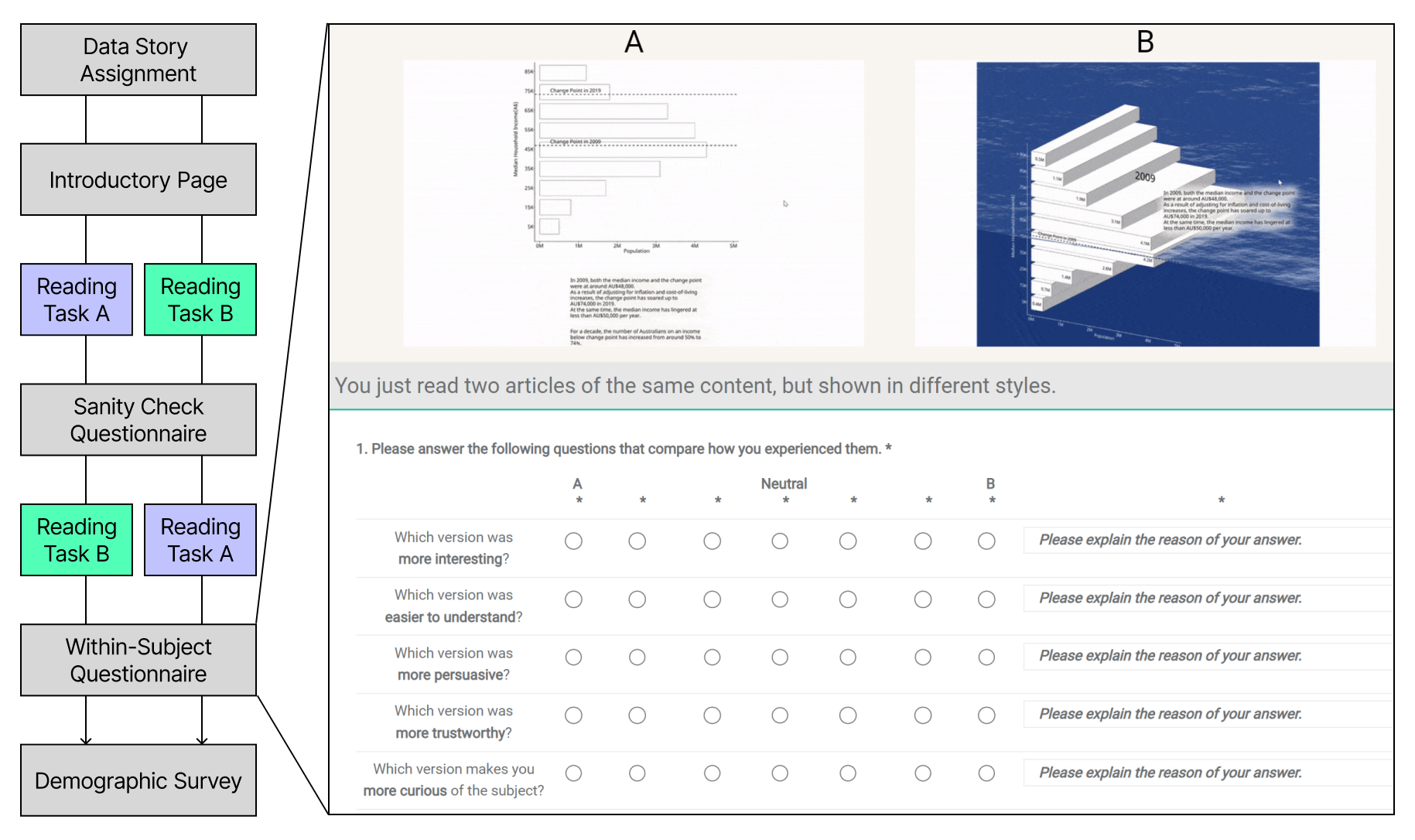}
    \caption{The within-subjects study procedure and the questionnaire page}
    \label{fig:procedure}
\end{figure*}

To answer the second research question, ``How does spatial immersion affect the reader’s experience?'', we chose five cognitive factors related to reading experiences of online data stories, and conducted a crowdsourced user study to collect quantitative and qualitative responses.

\subsection{Participants}
We recruited 600 participants from an online crowdsourcing platform \footnote{https://www.prolific.co/}, who accessed the study from UK~(436 participants), USA~(140), Canada~(9), Ireland~(8), and Australia~(7). Their ages range from 18 to 60~($\textit{Mean} = 36.9$, $\textit{SD} = 10.2$). Most of them had undergraduate degrees~($47.8\%$), followed by graduate degrees~($19.7 \%$), technical/community college~($18.3\%$), High school diploma/A-levels~($10.8\%$), and doctorate degree~($3.0 \%$). 
Participants who completed the entire study received a payment of GBP \pounds1.34.
In addition, the bonus criteria were mentioned at the beginning of the study to encourage sincere participation, and we paid a bonus GBP \pounds1.00 to those who provided at least one full reasoning sentence for each factor.

\subsection{Procedure}
Each participant is randomly assigned to one of the four data stories and a pair of the three conditions~(e.g., {\static}-{\animated}, {\animated}-{\immersive}, or {\immersive}-{\static}). 
The within-subject study procedure is illustrated in Figure~\ref{fig:procedure}. First, the landing page explains the study procedure and prohibited actions (e.g., refreshing the browser, leaving the page, taking a long break). Participants were also advised to adjust the size or zoom level of their browser windows to ensure they could read the articles correctly. The participants then read the first condition and answered the three questions for a knowledge check. To minimize potential ordering effects and learning biases, the order of the two conditions is counterbalanced across participants. Subsequently, participants read the second condition and proceeded to the questionnaire page, where they were asked to give relative scores of the two conditions on a 7-point Likert scale for the five evaluation metrics, described below, with short comments on their reasons for each score. Finally, participants completed a standard demographic survey. There was no time restriction during the entire process. 

\subsection{Evaluation Metrics} \label{sec:evaluationMetrics}  
Visual data storytelling has a wide range of effects on the reading experience \cite{Figueiras2014}. Researchers have conducted empirical studies on the engagement \cite{Mckenna2017}, likeability \cite{Figueiras2014}, comprehension \cite{Figueiras2014}, memorability \cite{hullman2013deeper, borkin2013makes}, and persuasiveness \cite{pandey2014persuasive, muehlenhaus2012if, liem2020structure} of visual data stories, regardless of how immersive they are. However, there are few (if any) empirical studies that focus on the (both positive and negative) effects of immersion in the context of visual data storytelling \cite{isenberg_immersive_2018}. Therefore, although we wanted to test as many metrics as possible, we decided to concentrate on the five criteria that are relevant to our research goals and suitable for our study setting, as outlined below. 

The first two metrics, \emph{interest} and \emph{curiosity}, are chosen to assess the level of engagement offered by the data story \cite{isenberg_immersive_2018, hidi2019interest} and to test whether spatial immersion makes visual data stories significantly more engaging.  
The first metric, \emph{interest}, focuses on immediate engagement that can be evoked by various aspects, such as color, animation, topic, and interactivity. In contrast, \emph{curiosity} pinpoints the longer and deeper motivation of the reader to better understand the topic \cite{hidi2019interest}.

We also considered metrics to evaluate the effectiveness of immersive data stories to reinforce, change, or shape the reader's attitudes \cite{Kukkonen2009}. 
For example, data stories can be designed to make readers feel specific emotions (e.g., anger, sadness, surprise, and joy), or to influence the reader's attitude towards the topic \cite{pandey2014persuasive}. However, the four data stories in the study were designed to objectively describe and explain the phenomenon rather than to change the reader's emotions or attitudes. Hence, we asked participants to compare a pair of data stories for their \emph{persuasiveness}, instead of asking if their pre-existing emotions or attitudes had changed during the reading. Furthermore, we chose to include \emph{trustworthiness}, a quality that enhances the credibility of persuasive systems \cite{Kukkonen2009}, since we were ambivalent about whether rich and dynamic visual effects of immersive stories would make the message more truthful, fair and unbiased, or if they would have the opposite effect.

Visual data stories are frequently used to facilitate the understanding and memorization of the content of the story \cite{shi2020}. However, we were uncertain whether immersion would further enhance the benefit, since data visualization experts have told that ornamental visuals and 3D graphics can distort visual perception and consequently hinder comprehension \cite{Zacks1998ReadingBG, skau2015evaluation}. Therefore, we asked the participants to evaluate which variant was \emph{easier to understand}. 
We decided to exclude some relevant metrics (e.g., memorability) that are not suitable for the study setting. For example, each participant was exposed to a pair of different variants during the study. This repeated exposure can cause carryover effects, making it difficult to measure memorability accurately.  

\subsection{Quantitative Analysis \& Result}

\begin{figure*}[t]
    \centering
    \includegraphics[width=0.98\textwidth,keepaspectratio]{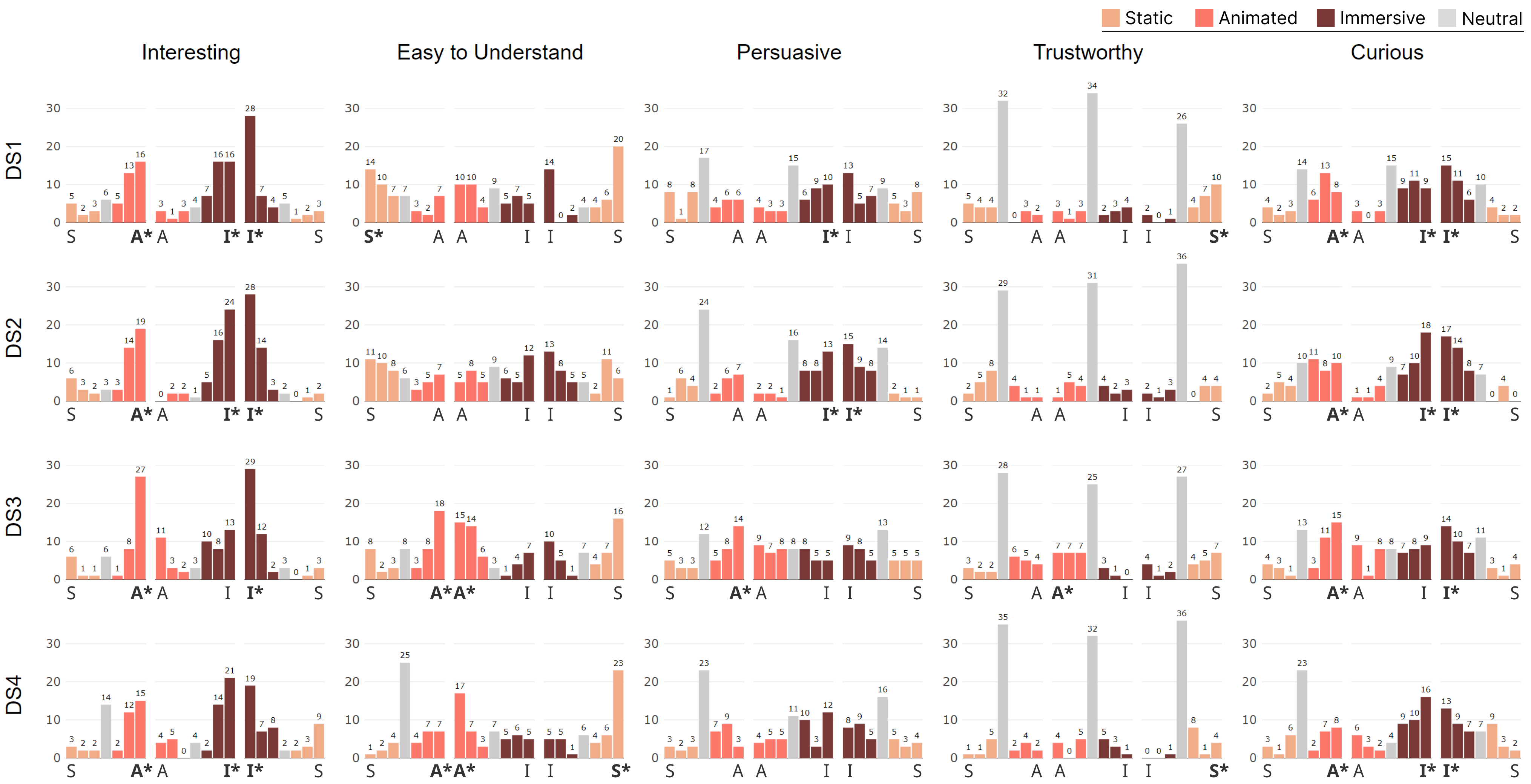}
    \caption{A group of frequency bar charts showing participant ratings on the cognitive factors for each story. We linearly concatenated bar charts of three comparison groups: S-A, A-I, and I-S, where S, A, and I stand for \static, \animated, and \immersive\ respectively. Gray indicates neutral scores. For each bar chart, the one-sample Wilcoxon signed-rank test was conducted to determine whether the median was significantly different from the neutral value. The variant name was highlighted with an asterisk if the results indicated that the median was significantly biased toward that variant.}
    \vspace{-6pt}
    \label {fig:histogram}
\end{figure*}

As participants compared two assigned variants for the five cognitive factors on a 7-point Likert scale, the histograms in Figure \ref{fig:histogram} visualize the number of votes for each variant. For each of the 60 comparisons, we evaluated the potential biases toward specific variants using Wilcoxon's one-sample signed-rank test and highlighted statistically significant biases ($p < 0.05$) with * symbols. The following paragraphs summarize the findings of each factor. 

\paragraphHeading{Interest} is the factor at which the \immersive\ stories were the most effective, and the \animated\ was more interesting than the \static.
In particular, the \immersive\ ones in DS1, DS2, and DS4 were significantly more interesting than the others. More than half of the participants in DS1, DS2, and DS3 rated extremely positive to the \immersive\ stories when compared to the \static.

\paragraphHeading{Ease of Understanding} is the factor where the \static\ and \animated\ stories were more effective than the \immersive. In specific, the \static\ ones received the largest number of votes for DS1. Also, for DS4, the \static\ outperformed the \immersive, receiving 33 out of 50 votes. The \animated\ stories were perceived as significantly easier to understand than the other variants for DS3 and DS4. 

\paragraphHeading{Persuasiveness} is the factor that no variant consistently outperformed the others. 
For DS1 and DS2, the \immersive\ condition was perceived to be more persuasive than the \animated\ and \static. However, in DS3, the \animated\ performed better than the others for this factor.

\paragraphHeading{Trustworthiness} is the factor that the largest ratio~$(371/600$, $61.8\%)$ of participants did not choose any winning variant. However, the \immersive\ stories were less trustworthy than the \animated\ for DS3 and the \static\ for DS1 and DS4.

\paragraphHeading{Curiosity} is the factor that shows a pattern similar to \emph{interest}. 
For DS1, DS2, and DS4, participants became more curious about the subject when reading the \immersive\ stories. 
Across all the stories, the \animated\ ones were more effective at making the participants curious about the subject than the \static. 

\subsection{Qualitative Analysis \& Result} \label{subsection:qualitativeresult}
While doing the questionnaire, the participants wrote short comments about their reasoning for each answer. We conducted an iterative open-coding process on them separately for the three experimental conditions. 
During the first round of open coding, we found that most of the comments contain which feature of the data story triggered specific emotions. 
For each experimental condition, we generated the initial codes employing such emotion and perceived features.
We then clustered them until we got five to six mid-level categories per condition. Finally, we identified four higher-level topics: (1) visual engagement, (2) reading cost, (3) trust in communication, and (4) informative guidance.

\paragraphHeading{Visual Engagement.} \label{subsubsection:visualengagement}
Engagement derived from visual elements is a frequently discussed topic. In particular, the participants stressed that animated transitions increased their interest and curiosity. Also, 3D charts were engaging to interact with and looked fancy in its visual properties such as color and texture.

\begin{itemize}
   
    \item \textbf{Interacting with animated graphs in 3D is engaging}~$(27/40$, $68\%)$ is the most frequently mentioned positive impact of the \immersive\ stories. Twelve participants felt engaged while exploring the 3D space with mouse interaction, and four even wanted to explore the visualization further, saying ``It had me more interested to learn what each axis was and why the graph was presented that way.'' On the other hand, 14 participants simply enjoyed watching the 3D models and animations, describing them ``higher in quality'', ``informative'', ``cinematic'', and ``visually appealing.'' As another kind of engagement, five participants felt involved in the situation, such as ``It immersed me into the experience, such as being on a roller coaster of the stock market trend.''

    \item \textbf{Fancy 3D visual is engaging and immersive}~$(25/40$, $63\%)$ explains why some participants preferred the \immersive\ ones over the \animated. The participants commented that they paid more attention because color, texture, animation, and 3D effects looked fancy and attractive. They also described 3D visuals as enjoyable, engaging, by saying, ``makes me more curious to keep seeing the animations and reading the text,'' and immersive, with a comment like ``It was almost an adventure!'' Seven participants felt as if they were exploring the data in 3D, mentioning they felt more ``drawn in'' and described the experience as being ``up and close, personally looking into the statistics of the data.''

    \item \textbf{Animated scrolly interaction is engaging}~$(15/40$, $38\%)$ is the reason why participants perceived the \animated\ stories interesting, making curious about the topic. They describe the \animated\ ones as ``visually appealing'', ``modern'', ``attention-grabbing'', ``novel'', and ``higher in quality''. Five participants found the animation with rich visuals engaging. The other five wrote that the scrollytelling format made the \animated\ interactive. Three participants described their involvement as feeling like they were ``inside'' the data.

\end{itemize}

\paragraphHeading{Reading Cost.} \label{subsubsection:readingcost}
Some participants found the \static\ stories clearer and easier to read, since they employ the classical structure of articles on the Web. In contrast, they complained that the \animated\ and \immersive\ stories demand more time and effort to read. Likewise, some participants found the \animated\ was easier to read, compared to the \immersive\ ones.  

\begin{itemize}
    \item \textbf{Stationary graphic is easier to understand}~$(16/40$, $40\%)$ is why the \animated\ was more preferred than the \immersive. Graphics in the \animated\ stories were less dynamic because they used 2D, static camera angle, and a fixed canvas. The participants described that less dynamicity made it easier to focus, read, and interpret. At the same time, they mentioned that 3D graphs, animation, and interactivity caused distraction and disorientation. 

    \item \textbf{Traditional 2D graph is quicker to understand}~$(16/40$, $40\%)$ contributed to the participant’s preference for the \static. The participants perceived the \static\ stories as simpler than the \immersive\ in their reading. Most of the participants~(14) described that the simplicity made it quicker and easier to read, follow, and navigate. Seven of them noted that it also increased interest, curiosity, and persuasion, as one mentioned, ``If I am easily able to find out the data, then I'm more likely to pay attention.'' Other four responses mentioned that the simpler design felt professional, traditional, and familiar. 
    
    \item \textbf{Simpler presentation is clearer to read}~$(11/40$, $28\%)$ is the main reason why some participants preferred the \static\ more than the \animated. The participants reported that the \static\ had a relatively ``plain visual'' and ``simple layout'', and offered a ``neater presentation''. Six participants preferred straightforward flow, saying ``I would side more with the \emph{static}\footnote{The italicized \emph{static}, \emph{animated}, and \emph{immersive} in the quotations were substituted by the authors to denote the variants.} since it gets to the point without excessively drawing out the content.'' Also, five participants pointed out that reading the article with static text chunks and graphs was easier on their eyes.

    \item \textbf{3D movements are hard to follow}~$(11/40$, $28\%)$ is a negative factor toward the \immersive\ stories. Some participants described being interfered with by ``scrolly interactive'', ``excessively moving'', or ``flashy'' elements. The participants felt it was difficult to read and follow because of the data representation's flashy movement. Five participants mentioned that they could concentrate more on the information in the \static, while feeling distracted by moving graphs in the \immersive\ ones. One comment pointed out that moving graphs grabbed attention, however, it was harder to see for a longer time.

    \item \textbf{High cost of using scrolly interaction}~$(6/40$, $15\%)$ had negative impacts on participants' reactions toward the \animated\ stories. Two participants found scrolly interaction tedious, because they had to wait until animated transitions finished. Three other participants preferred the \static\ ones, mentioning ``I chose the \emph{static} because I don't have to keep scrolling down for an explanation.''

\end{itemize}

\paragraphHeading{Trust in Communication.} \label{subsubsection:trustincom} Participants often felt suspicious about the author's intentions. In particular, they found the excessive use of visuals in the \animated\ to be gimmicky and disruptive to their reading. On the other hand, the participants described the \immersive\ stories as deceptive but relatively convincing compared to the \static\ and more professionally made than the \animated.

\begin{itemize}
    \item \textbf{3D graphic seems professional but deceptive}~$(13/40$, $33\%)$ as six participants perceived the \immersive\ more professional, requiring more time and effort for production. However, seven participants were concerned about the trustworthiness of 3D charts, and said that 2D graphs would be more accurate. One of them mentioned, ``2D charts made it easy to understand what was going on. 3D charts look better, but they are deceptive.''
    
    \item \textbf{Too much visual can be gimmicky}~$(10/40$, $25\%)$ shows that the overly flashy, moving visuals of the \animated\ ones gave participants negative impacts. Six participants described difficulties in understanding the information. Four of them mentioned that moving graphics and words were distracting, and the other two participants complained that excessive visual effects made it difficult to find where to look at. Other six participants highlighted that they felt it was harder to interpret data or messages, which led to a decrease in persuasiveness and trustworthiness by saying, ``the \emph{animated} was too gimmicky to be trustworthy,'' and ``flashy presentation makes me question whether the presenter/author is trying to hide something.''

    \item \textbf{Fancy \& flashy visual is unreliable}~$(7/40$, $18\%)$ had extremely negative impacts on the \immersive. Three participants found that the 3D visuals, animation, and camera movements of the \immersive\ made it harder to compare and understand the data. Five commented that, while the \static\ stories simply set out the facts, the \immersive\ was trying to deceive the reader with a flashy, convincing presentation. Some even suspected the author's intention by saying, ``the \emph{immersive} looks like trying to distract you with fancy presentation like a second hand car dealer.''

\end{itemize}

\paragraphHeading{Informative Guidance.} \label{subsubsection:informativeguidances}
The participants mentioned that dynamic visual effects provide informative guidance to understand the narrative. For example, the fine-grained synchronization of visuals and texts of the \animated\ and \immersive\ stories made it easier to follow. In addition, the sense of 3D space in the \immersive\ helped them grasp the scale and trends of the data. Lastly, metaphoric visuals illustrated the author's message and helped follow the data stories.

\begin{itemize}
    
    \item \textbf{3D is easier to visualize the trends}~$(9/40$, $23\%)$ explains that the nature of the 3D in the \immersive\ stories helped understanding the data visualization. 
    In other words, properties such as width, height, and depth were able to visualize the data trends in an engaging way. 
    For example, three participants found that the height of the 3D bars in DS4 made it easier to grasp data scale, upward and downward trends, and high points, and two other participants described that value changes encoded using length made DS3 more persuasive. 
    Also, the participants mentioned that 3D models are something to interact with, as commented ``It gives you a ‘world’ to see the graph with,'' and ``you can physically see the data rather than it just being numbers in a bar chart.''
    
    \item \textbf{Bit-by-bit synchronization helps understanding}~$(8/40$, $20\%)$ is a major positive impact of the \animated\, as it helped the participants understand the data stories. They also pointed out that animations synchronized with the narrative flow were particularly useful. Four participants mentioned that gradually going into details with smaller chunks of information helped them focus on each bit of content, as one described, ``The information came bit by bit. It allowed me to digest the information easier and help me understand it better.'' Three participants also mentioned that seeing text tied up with visualization made the article easier to understand, saying ``The graph and text joined up together better, and just made more clear sense in my head.''
    
    \item \textbf{Real world metaphor is illustrative and involving}~$(4/40$, $10\%)$ is a small group of codes that explain how the \immersive\ stories effectively contextualize some participants. In particular, real world metaphors (i.e., the ocean and the roller-coaster rails) employed in DS2 and DS3 helped participants get involved into the narrative, by saying, ``the \emph{immersive} was more visual so easy to understand (e.g., water level rising).''

\end{itemize}

\section{Discussion}
\label{sec:discussion}

The user study showed that there is no single winning condition for all five cognitive factors. In Section \ref{sec:discussion-each-condition}, we discuss the strengths and weaknesses of the three design variants and conclude with insights on their usage. In Section \ref{sec:discussion-design_implication}, we discuss the design implications of when and how to develop immersive data stories. Lastly, Section \ref{sec:discussion-authoring_tool} discusses unmet needs and opportunities for authoring tools, including HTML5 and third-party libraries to produce charts and 3D environments. 

\subsection{Strengths, Weaknesses, and Usage Insights}
\label{sec:discussion-each-condition}

We designed the \static\ as the baseline condition, which does not evoke the spatial immersion. 
This \static\ design helped the participants read the story with less effort, thanks to the simple interactivity, stationary graphics, and familiar layout~(\S\ref{subsubsection:readingcost}). 
Meanwhile, they felt the \static\ stories more concise and straightforward, since it only covers essential information and interactivity. 
In particular, the participants appreciated compact layout featuring chunky texts and multiple charts as it does not force them to scroll down for an explanation.
As another advantage, the \static\ condition is perceived as highly trustworthy due to its tidiness~(\S\ref{subsubsection:trustincom}).
While most participants similarly trusted given conditions as they share the same data source and texts, the fact that the \static\ did not attempt to deceive the readers made it more reliable.
To sum up, the \static\ condition is the least engaging, but still a good option for casual readers who do not want to spend excessive time and effort for interpreting data visualization with complex styles.

We designed the \animated\ condition to incorporate chart animation effects similar to the \immersive\ condition but without evoking a sense of space.
The \animated\ charts and overlaid texts were still engaging to read the story~(\S\ref{subsubsection:visualengagement}), and they also evoked curiosity on the subjects efficiently.
Furthermore, the synchronized chart gradually explaining the story~(\S\ref{subsubsection:informativeguidances}) also made it comprehensible.
Despite several design patterns applied to chart and layout, the \animated\ were still simpler when compared to the \immersive\ as the animation was 2D and only played within chart.
This made the story clear, reliable, and easy to read, like the advantages of the \static.
Readers sometimes perceive that the movements of the charts and texts are excessive. 
However, there are numerous quality free examples online, which assure the solid user experience and reduce the production effort.
In conclusion, the \animated\ condition is a good option when authors desire a promising and efficient design for enhancing engagement without the risks of adopting state-of-the-art technology.

The \immersive\ condition used 3D animated graphics eliciting a sense of space.
These aesthetic features efficiently grabbed attention, and led more engaging experience~(\S\ref{subsubsection:visualengagement}).
For example, the 3D camera movements are cinematic and involving, giving adventurous experience inside a virtual space of data visualization.
When this animation demonstrates the story and provides a 3D sense of space, the participants also found it easier to visualize the data trends and follow the story~(\S\ref{subsubsection:informativeguidances}). 
On the other hand, some participants emphasized these fancy and flash visuals were distracting and unreliable~(\S\ref{subsubsection:trustincom}). 
In specific, we can expect the best performance in terms of \emph{interest} and \emph{curiosity}, however, we should be careful not to make the animation distracting and gimmicky. 

\subsection{Design Implications for Immersive Data Story}
\label{sec:discussion-design_implication}
Regarding the advantages and drawbacks of the \immersive\ condition, various methods exist to enhance the performance of the \immersive, such as amplifying strengths to overshadow weaknesses, or concealing the disadvantages and take the readers' attention to the others.
However, these ideas are still vague, thus we suggest two applicable design implications for making the \immersive\ story more preferrable.

\paragraphHeading{Organize the \immersive\ story with other designs.}
\label{sec:discussion-organize}
Given that the three conditions have distinct strengths in terms of five cognitive factors, we can design the story as an interplay of different designs.
By showing a clear 2D view in between 3D immersive scenes, readers may perceive the data story as understandable and trustworthy.
Conversely, they may become highly engaged with the immersive virtual space suddenly transitioned from a simple 2D scene.
S19 in Table \ref{table:visual data story cases} is a good example of an interplay. 
It begins with the \immersive\ chart and text layout, then simply lists up arguments using pairs of text chunk and \static\ charts. This combination can catch reader's attention to the topic in early stages, and end up with reliable and comprehensible representation to make the message clear.
Similarly, the \textit{Martini Glass Structure}\cite{Segel2010} also represents mixed scenes which begins with an author-driven flow, transitioning to a reader-driven stage for interactivity and exploration.
Given the participants' feedback showing increased time and focus on engaging visuals and interactivity, we anticipate that readers will likely end up with higher level of curiosity.

In order to create a seamless story utilizing multiple scenes, designing the story using \texttt{DP6}~(Dynamic Dimension) could be a practical choice.
As an example of using \texttt{DP6}, we made DS2 using the 3D bar chart that resembles an iceberg floating on the ocean surface.
This 3D bar chart has meaningless depth, which is designed only for giving a symbolic appearance.
Thus, we provided the only transition between the front view and 3D overview to efficiently convey its metaphorical message.
However, diverse transitions are available when \texttt{DP6} is applied to a 3D chart with three different variables.
In S6, the 3D patch visualization has three different combinations of axes with front, side, and top views.
These 2D charts and 3D overview are alternately displayed throughout the story, smoothly changing the dimension and the axes, which helps following the subsequent text and chart.

\paragraphHeading{Use the \immersive\ design when the data is easy to grasp.}
As previously discussed, the readers show increased focus and engagement on embellished graphics.
Also, the \immersive\ design might create a positive impression such as polished, tech savvy and futuristic, rather than a serious scientific article or a presentation given at school.
Thus, when readers perceive the story as easily comprehensible in any reasons, leveraging an \immersive\ design presents an opportunity to make such impacts with less disadvantages.
For example, both S5 and S8 use various design use color and height encoding on population variable to represent the data like a mountainous terrain. 
Also, S5 utilizes view transition from 2D to 3D and several views from angles, and S8 provides free exploration using mouse interaction.
However, both S5 and S8 remain easy to follow as their narratives are just exploring populations of cities and the visualizations only show the cities' locations and populations on the world map. 
Additionally, they allow readers to explore and identify both the plain areas indicating low population and the red, tall data points grouped like a mountain, without demanding any other difficult tasks.
This makes readers to have little stress on digesting the story and to focus on exploring the data of cities based on their interest, which presents an opportunity for employing \emph{immersive} designs.

\subsection{Authoring Tools for Immersive Data Story on the Web}
\label{sec:discussion-authoring_tool}
Even if practitioners had a good understanding of when and how to employ the design patterns, they are required to create and control text, charts, and other elements using general-purpose tools and grammars (e.g., CSS, D3, or Three.js). Due to the limited availability of high-level tools and the extensive amount of manual tasks, creating immersive data stories from scratch could be tedious and time-consuming. Based on our own experience in developing the four data stories, creating an \immersive\ story accounted for approximately 90 percent of the total development time. In the following, we discuss three domains of tools, which are (1) HTML and CSS standards, (2) tools for chart production, and (3) tools for creating 3D content, in terms of their roles in web data storytelling and how to enhance them. 

First, HTML and CSS standards are key building blocks that not only define the core narrative structure but also integrate other components (e.g., 2D/3D charts and visuals) into a meaningful way. In addition, practitioners use HTML to monitor interaction events, such as clicking buttons or scrolling pages, and trigger dynamic visual effects to evoke a sense of spatial immersion. For example, immersive data stories often detect the page scroll event and move elements at different speeds so that the readers would feel as if the moving elements were placed at varying depths on the z-axis. Although the parallax effect is a common use case of \texttt{DP5}, HTML and CSS standards currently do not offer a high-level grammar to use it, and thus authors have to specify desired visual effects from scratch. Likewise, if practitioners want to employ dynamic shadow effects, they will need to manually set the formula to calculate the direction and the size of the shadow. Therefore, to increase the efficiency of such tasks, we suggest that future standards of HTML and CSS can offer declarative controls over effects for spatial immersion, linked to z-depths of individual elements. Furthermore, it might also be beneficial to offer an easy-to-use high-level grammar to control basic rendering capabilities, such as light source, camera, and texture controls. With the above features, HTML and CSS would become an effective tool for authoring immersive web data stories on the Web.   

Second, although many JS chart libraries (e.g., Plotly, D3) support creating 3D charts, we did not use them to author the DSs, due to the lack of controls and features for immersive storytelling. For example, existing libraries do not offer an easy way to employ a variety of camera shots and transitions (\texttt{DP1}) or to switch between 2D and 3D camera modes (\texttt{DP6}). Moreover, existing libraries are designed to create standalone charts so that it is difficult (if possible) to pair tightly with the narrative flow. In consequence, we had to use Three.js, a general-purpose API for 3D Web graphics, to create and control the chart graphics from scratch.  

Another opportunity for chart libraries is to facilitate richer and realistic visualization. For example, although chart libraries visualize data using abstract geometries (e.g., rectangles, circles, and lines), immersive data stories often need to simulate a realistic appearance (\texttt{DP5}) via visual embellishments, such as pictograms \cite{Haroz_2015_isotype} or simplified / realistic representations of people \cite{morais2020showing}. Likewise, immersive data stories often create the feeling of virtual space by using smoothly and naturally moving elements, as described in \texttt{DP3}. For example, if a future library allows the author to encode markers as blood cells and to apply built-in physics engines, the story will be able to engage readers with the feeling of being immersed into the context. 

Lastly, most of the articles collected during the literature study, as well as the four data stories used in our experiment, employ Three.js\footnote{https://threejs.org/}, a popular JavaScript library to create and display 3D web graphics using WebGL, due to its simplicity (compared to WebGL), extensive features, and flexibility to create a wide range of 3D applications. However, using it for immersive data storytelling requires a comprehensive understanding of 3D graphics, including geometry, lighting, camera controls, and even custom shaders. Moreover, it does not provide built-in features for handling rich user interactions (e.g., zooming or filtering data), controlling cameras and transitions, and synchronizing them with the narrative flow. To allow non-programmers to create immersive data stories with fewer lines of code, Three.js (or other 3D libraries) in the future could include built-in high-level constructs for data storytelling (e.g., 3D charts, scatterplots, maps, network graphs, and timeline) as well as animation libraries for camera movements, object transformation, and scene transitions. Taking it further, a visual editor will be able to allow practitioners without programming experience to create their own immersive data stories.

\subsection{Limitations and Future Work}
As described in Section \ref{sec:evaluationMetrics}, we focused on the five cognitive factors that are relevant and suitable for the study setting. However, future work may employ other metrics as well. For example, examining emotional impacts such as surprise, empathy, and seriousness could provide more insight into immersive data storytelling \cite{lan2021kineticharts}. Future work could measure the degree of immersion by gathering eye movements or task completion time after immersive experience\cite{Jennett2008}.

Since we performed the literature review in the initial phase of the project, every article in our collection was published before 2022. Assuming that the continuous production of data stories might invoke spatial immersion from various sources on the Web, an exhaustive and comprehensive survey of the latest data stories could be beneficial to investigate the current trends and techniques in this area\cite{stolper_emerging_2016}.

Existing web-based data stories have already shown their potential to employ interactions for both 2D displays and VR.
For example, ``Is the NASDAQ in Another Bubble?''\footnote{https://graphics.wsj.com/3d-nasdaq/} allows readers to select between these two formats.
Similarly, ``Passage of Water''\footnote{https://artsandculture.google.com/experiment/passage-of-water/dAElpEyEjuE9XQ} integrates a scrollytelling structure with VR-like free exploration.
Therefore, we anticipate that research focused on multi-device compatibility with a single narrative is likely to be conducted in the near future.
This seems feasible with current web graphic technology, since the 3D rendering library used in our development named Three.js already offered APIs for building VR content\footnote{https://threejs.org/docs/\#manual/ko/introduction/How-to-create-VR-content}.
Otherwise, future work can use the camera and data visualization control algorithms used in authoring tools for 3D data visualizations\cite{ren2019visualization, cordeil2017imaxes, sicat2018dxr}. 
A research field of immersive data analytics has developed various user interfaces, visual encoding, and design techniques.
If such researches are integrated with concepts of spatial immersion, automatic scene generation\cite{evin2022cine, huang2019learning} and camera techniques\cite{he2023virtual}, it seems evident that web data stories compatible with both 2D display and VR context would be efficiently created.
\section{Conclusion}
As technology for spatial immersion advances, we will see an increasing number of immersive data stories. In this paper, we identify six design patterns from a literature review of existing visual data stories on the Web. We also developed the four data stories with three design variations. The results of our crowd-sourced user study show that the \immersive\ condition outperformed the \animated\ and the \static\ conditions for three of the five cognitive factors. We believe that our discussions would be helpful for developers and researchers working on creating immersive data storytelling. 





\bibliographystyle{IEEEtran}
\bibliography{refs}

\begin{thebibliography}{10}
\providecommand{\url}[1]{#1}
\csname url@samestyle\endcsname
\providecommand{\newblock}{\relax}
\providecommand{\bibinfo}[2]{#2}
\providecommand{\BIBentrySTDinterwordspacing}{\spaceskip=0pt\relax}
\providecommand{\BIBentryALTinterwordstretchfactor}{4}
\providecommand{\BIBentryALTinterwordspacing}{\spaceskip=\fontdimen2\font plus
\BIBentryALTinterwordstretchfactor\fontdimen3\font minus \fontdimen4\font\relax}
\providecommand{\BIBforeignlanguage}[2]{{%
\expandafter\ifx\csname l@#1\endcsname\relax
\typeout{** WARNING: IEEEtran.bst: No hyphenation pattern has been}%
\typeout{** loaded for the language `#1'. Using the pattern for}%
\typeout{** the default language instead.}%
\else
\language=\csname l@#1\endcsname
\fi
#2}}
\providecommand{\BIBdecl}{\relax}
\BIBdecl

\bibitem{ermi2007analyzing}
L.~Ermi and F.~M{\"a}yr{\"a}, ``Analyzing immersion,'' \emph{Worlds in play: International perspectives on digital games research}, vol.~21, p.~37, 2007.

\bibitem{mcmahan2013immersion}
A.~McMahan, ``Immersion, engagement, and presence: A method for analyzing 3-d video games,'' in \emph{The video game theory reader}.\hskip 1em plus 0.5em minus 0.4em\relax Routledge, 2013, pp. 67--86.

\bibitem{ryan2015narrative}
M.-L. Ryan, \emph{Narrative as virtual reality 2: Revisiting immersion and interactivity in literature and electronic media}.\hskip 1em plus 0.5em minus 0.4em\relax JHU press, 2015.

\bibitem{Witmer1998}
\BIBentryALTinterwordspacing
B.~G. Witmer and M.~J. Singer, ``{Measuring Presence in Virtual Environments: A Presence Questionnaire},'' \emph{Presence: Teleoperators and Virtual Environments}, vol.~7, no.~3, pp. 225--240, 06 1998. [Online]. Available: \url{https://doi.org/10.1162/105474698565686}
\BIBentrySTDinterwordspacing

\bibitem{murray1997hamlet}
\BIBentryALTinterwordspacing
J.~Murray, \emph{Hamlet on the Holodeck: The Future of Narrative in Cyberspace}.\hskip 1em plus 0.5em minus 0.4em\relax Free Press, 1997. [Online]. Available: \url{https://books.google.co.kr/books?id=bzmSLtnMZJsC}
\BIBentrySTDinterwordspacing

\bibitem{slater1997framework}
M.~Slater and S.~Wilbur, ``A framework for immersive virtual environments (five): Speculations on the role of presence in virtual environments,'' \emph{Presence: Teleoperators \& Virtual Environments}, vol.~6, no.~6, pp. 603--616, 1997.

\bibitem{ROONEY2012405}
\BIBentryALTinterwordspacing
B.~Rooney, C.~Benson, and E.~Hennessy, ``The apparent reality of movies and emotional arousal: A study using physiological and self-report measures,'' \emph{Poetics}, vol.~40, no.~5, pp. 405--422, 2012. [Online]. Available: \url{https://www.sciencedirect.com/science/article/pii/S0304422X12000502}
\BIBentrySTDinterwordspacing

\bibitem{nilsson2016immersion}
N.~C. Nilsson, R.~Nordahl, and S.~Serafin, ``Immersion revisited: A review of existing definitions of immersion and their relation to different theories of presence,'' \emph{Human technology}, vol.~12, no.~2, pp. 108--134, 2016.

\bibitem{slater2003}
M.~Slater, ``A note on presence terminology,'' \emph{Presence Connect}, vol.~3, 01 2003.

\bibitem{Zhi2019}
Q.~Zhi, A.~Ottley, and R.~Metoyer, ``Linking and layout: Exploring the integration of text and visualization in storytelling,'' in \emph{Computer Graphics Forum}, vol.~38, no.~3.\hskip 1em plus 0.5em minus 0.4em\relax Wiley Online Library, 2019, pp. 675--685.

\bibitem{Mckenna2017}
S.~McKenna, N.~Henry~Riche, B.~Lee, J.~Boy, and M.~Meyer, ``Visual narrative flow: Exploring factors shaping data visualization story reading experiences,'' in \emph{Computer Graphics Forum}, vol.~36, no.~3.\hskip 1em plus 0.5em minus 0.4em\relax Wiley Online Library, 2017, pp. 377--387.

\bibitem{Zhang2017}
C.~Zhang, A.~Perkis, and S.~Arndt, ``Spatial immersion versus emotional immersion, which is more immersive?'' in \emph{2017 Ninth International Conference on Quality of Multimedia Experience (QoMEX)}, 2017, pp. 1--6.

\bibitem{Chen2016}
X.~Chen, J.~Z. Self, L.~House, and C.~North, ``Be the data: A new approach for lmmersive analytics,'' in \emph{2016 Workshop on Immersive Analytics (IA)}, 2016, pp. 32--37.

\bibitem{Willett2016}
W.~Willett, Y.~Jansen, and P.~Dragicevic, ``Embedded data representations,'' \emph{IEEE Transactions on Visualization and Computer Graphics}, vol.~23, no.~1, pp. 461--470, 2017.

\bibitem{klemm_interactive_2014}
P.~Klemm, S.~Oeltze-Jafra, K.~Lawonn, K.~Hegenscheid, H.~Völzke, and B.~Preim, ``\BIBforeignlanguage{eng}{Interactive {Visual} {Analysis} of {Image}-{Centric} {Cohort} {Study} {Data}},'' \emph{\BIBforeignlanguage{eng}{IEEE transactions on visualization and computer graphics}}, vol.~20, no.~12, pp. 1673--1682, Dec. 2014.

\bibitem{bezerianos_perception_2012}
A.~Bezerianos and P.~Isenberg, ``\BIBforeignlanguage{eng}{Perception of {Visual} {Variables} on {Tiled} {Wall}-{Sized} {Displays} for {Information} {Visualization} {Applications}},'' \emph{\BIBforeignlanguage{eng}{IEEE transactions on visualization and computer graphics}}, vol.~18, no.~12, pp. 2516--2525, Dec. 2012.

\bibitem{Lee2015}
B.~Lee, N.~H. Riche, P.~Isenberg, and S.~Carpendale, ``More than telling a story: Transforming data into visually shared stories,'' \emph{IEEE computer graphics and applications}, vol.~35, no.~5, pp. 84--90, 2015.

\bibitem{Segel2010}
E.~Segel and J.~Heer, ``Narrative visualization: Telling stories with data,'' \emph{IEEE Transactions on Visualization and Computer Graphics}, vol.~16, no.~6, pp. 1139--1148, 2010.

\bibitem{dove2012narrative}
G.~Dove and S.~Jones, ``Narrative visualization: Sharing insights into complex data,'' 2012.

\bibitem{hullman2011visualization}
J.~Hullman and N.~Diakopoulos, ``Visualization rhetoric: Framing effects in narrative visualization,'' \emph{IEEE transactions on visualization and computer graphics}, vol.~17, no.~12, pp. 2231--2240, 2011.

\bibitem{Amini2015}
F.~Amini, N.~Henry~Riche, B.~Lee, C.~Hurter, and P.~Irani, ``Understanding data videos: Looking at narrative visualization through the cinematography lens,'' in \emph{Proceedings of the 33rd Annual ACM conference on human factors in computing systems}, 2015, pp. 1459--1468.

\bibitem{Csikszentmihalyi2014}
M.~Csikszentmihalyi, J.~Nakamura, and M.~Csikszentmihalyi, ``The concept of flow,'' \emph{Flow and the foundations of positive psychology: The collected works of Mihaly Csikszentmihalyi}, pp. 239--263, 2014.

\bibitem{Agarwal2000}
R.~Agarwal and E.~Karahanna, ``Time flies when you're having fun: Cognitive absorption and beliefs about information technology usage,'' \emph{MIS quarterly}, pp. 665--694, 2000.

\bibitem{Brown2004}
E.~Brown and P.~Cairns, ``A grounded investigation of game immersion,'' in \emph{CHI'04 extended abstracts on Human factors in computing systems}, 2004, pp. 1297--1300.

\bibitem{coomans1997towards}
M.~K. Coomans and H.~J. Timmermans, ``Towards a taxonomy of virtual reality user interfaces,'' in \emph{Proceedings. 1997 IEEE Conference on Information Visualization (Cat. No. 97TB100165)}.\hskip 1em plus 0.5em minus 0.4em\relax IEEE, 1997, pp. 279--284.

\bibitem{curran2018factors}
N.~Curran, ``Factors of immersion,'' \emph{The Wiley Handbook of Human Computer Interaction}, vol.~1, pp. 239--254, 2018.

\bibitem{o2010development}
H.~L. O'Brien and E.~G. Toms, ``The development and evaluation of a survey to measure user engagement,'' \emph{Journal of the American Society for Information Science and Technology}, vol.~61, no.~1, pp. 50--69, 2010.

\bibitem{adams2006fundamentals}
E.~Adams and A.~Rollings, \emph{Fundamentals of game design (game design and development series)}.\hskip 1em plus 0.5em minus 0.4em\relax Prentice-Hall, Inc., 2006.

\bibitem{bach2018narrative}
B.~Bach, M.~Stefaner, J.~Boy, S.~Drucker, L.~Bartram, J.~Wood, P.~Ciuccarelli, Y.~Engelhardt, U.~Koeppen, and B.~Tversky, ``Narrative design patterns for data-driven storytelling,'' in \emph{Data-driven storytelling}.\hskip 1em plus 0.5em minus 0.4em\relax AK Peters/CRC Press, 2018, pp. 107--133.

\bibitem{Conlen2023}
M.~Conlen, J.~Heer, H.~Mushkin, and S.~Davidoff, ``Cinematic techniques in narrative visualization,'' \emph{arXiv preprint arXiv:2301.03109}, 2023.

\bibitem{shi2021communicating}
Y.~Shi, X.~Lan, J.~Li, Z.~Li, and N.~Cao, ``Communicating with motion: A design space for animated visual narratives in data videos,'' in \emph{Proceedings of the 2021 CHI conference on human factors in computing systems}, 2021, pp. 1--13.

\bibitem{lan2021kineticharts}
X.~Lan, Y.~Shi, Y.~Wu, X.~Jiao, and N.~Cao, ``Kineticharts: Augmenting affective expressiveness of charts in data stories with animation design,'' \emph{IEEE Transactions on Visualization and Computer Graphics}, vol.~28, no.~1, pp. 933--943, 2021.

\bibitem{xu2022wow}
X.~Xu, L.~Yang, D.~Yip, M.~Fan, Z.~Wei, and H.~Qu, ``From ‘wow’to ‘why’: Guidelines for creating the opening of a data video with cinematic styles,'' in \emph{Proceedings of the 2022 CHI Conference on Human Factors in Computing Systems}, 2022, pp. 1--20.

\bibitem{mayer2023characterization}
B.~Mayer, N.~Steinhauer, B.~Preim, and M.~Meuschke, ``A characterization of interactive visual data stories with a spatio-temporal context,'' in \emph{Computer Graphics Forum}, vol.~42, no.~6.\hskip 1em plus 0.5em minus 0.4em\relax Wiley Online Library, 2023, p. e14922.

\bibitem{tufte_1986}
E.~R. Tufte, \emph{The Visual Display of Quantitative Information}.\hskip 1em plus 0.5em minus 0.4em\relax USA: Graphics Press, 1986.

\bibitem{Bateman2010}
S.~Bateman, R.~L. Mandryk, C.~Gutwin, A.~Genest, D.~McDine, and C.~Brooks, ``Useful junk? the effects of visual embellishment on comprehension and memorability of charts,'' in \emph{Proceedings of the SIGCHI conference on human factors in computing systems}, 2010, pp. 2573--2582.

\bibitem{Boy2015_does_it_engage}
\BIBentryALTinterwordspacing
J.~Boy, F.~Detienne, and J.-D. Fekete, ``Storytelling in information visualizations: Does it engage users to explore data?'' in \emph{Proceedings of the 33rd Annual ACM Conference on Human Factors in Computing Systems}, ser. CHI '15.\hskip 1em plus 0.5em minus 0.4em\relax New York, NY, USA: Association for Computing Machinery, 2015, p. 1449–1458. [Online]. Available: \url{https://doi.org/10.1145/2702123.2702452}
\BIBentrySTDinterwordspacing

\bibitem{isenberg_immersive_2018}
\BIBentryALTinterwordspacing
P.~Isenberg, B.~Lee, H.~Qu, and M.~Cordeil, ``Immersive {Visual} {Data} {Stories},'' in \emph{Immersive {Analytics}}, K.~Marriott, F.~Schreiber, T.~Dwyer, K.~Klein, N.~H. Riche, T.~Itoh, W.~Stuerzlinger, and B.~H. Thomas, Eds.\hskip 1em plus 0.5em minus 0.4em\relax Cham: Springer International Publishing, 2018, pp. 165--184. [Online]. Available: \url{https://doi.org/10.1007/978-3-030-01388-2_6}
\BIBentrySTDinterwordspacing

\bibitem{Barral2020}
O.~Barral, S.~Lall{\'e}, and C.~Conati, ``Understanding the effectiveness of adaptive guidance for narrative visualization: a gaze-based analysis,'' in \emph{Proceedings of the 25th international conference on intelligent user interfaces}, 2020, pp. 1--9.

\bibitem{romat2020dear}
H.~Romat, N.~Henry~Riche, C.~Hurter, S.~Drucker, F.~Amini, and K.~Hinckley, ``Dear pictograph: Investigating the role of personalization and immersion for consuming and enjoying visualizations,'' in \emph{Proceedings of the 2020 CHI Conference on Human Factors in Computing Systems}, 2020, pp. 1--13.

\bibitem{crowdsourcing_user_studies}
\BIBentryALTinterwordspacing
A.~Kittur, E.~H. Chi, and B.~Suh, ``Crowdsourcing user studies with mechanical turk,'' in \emph{Proceedings of the SIGCHI Conference on Human Factors in Computing Systems}, ser. CHI '08.\hskip 1em plus 0.5em minus 0.4em\relax New York, NY, USA: Association for Computing Machinery, 2008, p. 453–456. [Online]. Available: \url{https://doi.org/10.1145/1357054.1357127}
\BIBentrySTDinterwordspacing

\bibitem{victor2011explorable}
B.~Victor, ``Explorable explanations,'' \emph{Online. http://worrydream. com/ExplorableExplanations}, vol.~17, 2011.

\bibitem{danchilla2012three}
B.~Danchilla and B.~Danchilla, ``Three. js framework,'' \emph{Beginning WebGL for HTML5}, pp. 173--203, 2012.

\bibitem{bostock2011d3}
M.~Bostock, V.~Ogievetsky, and J.~Heer, ``D$^3$ data-driven documents,'' \emph{IEEE transactions on visualization and computer graphics}, vol.~17, no.~12, pp. 2301--2309, 2011.

\bibitem{satyanarayan2016vega}
A.~Satyanarayan, D.~Moritz, K.~Wongsuphasawat, and J.~Heer, ``Vega-lite: A grammar of interactive graphics,'' \emph{IEEE transactions on visualization and computer graphics}, vol.~23, no.~1, pp. 341--350, 2016.

\bibitem{scrollytelling_2018}
D.~Seyser and M.~Zeiller, ``Scrollytelling – an analysis of visual storytelling in online journalism,'' in \emph{2018 22nd International Conference Information Visualisation (IV)}, 2018, pp. 401--406.

\bibitem{wolf_longforms_journalism_2016}
\BIBentryALTinterwordspacing
C.~Wolf and A.~Godulla, ``Potentials of digital longforms in journalism. a survey among mobile internet users about the relevance of online devices, internet-specific qualities, and modes of payment,'' \emph{Journal of Media Business Studies}, vol.~13, no.~4, pp. 199--221, 2016. [Online]. Available: \url{https://doi.org/10.1080/16522354.2016.1184922}
\BIBentrySTDinterwordspacing

\bibitem{Seyser_2018}
D.~Seyser and M.~Zeiller, ``Scrollytelling – an analysis of visual storytelling in online journalism,'' in \emph{2018 22nd International Conference Information Visualisation (IV)}, 2018, pp. 401--406.

\bibitem{sultanum2021leveraging}
N.~Sultanum, F.~Chevalier, Z.~Bylinskii, and Z.~Liu, ``Leveraging text-chart links to support authoring of data-driven articles with vizflow,'' in \emph{Proceedings of the 2021 CHI Conference on Human Factors in Computing Systems}, 2021, pp. 1--17.

\bibitem{vallandingham_so_nodate}
\BIBentryALTinterwordspacing
J.~Vallandingham, ``\BIBforeignlanguage{en}{So {You} {Think} {You} {Can} {Scroll} {Talk}}.'' [Online]. Available: \url{https://vallandingham.me/think_you_can_scroll.html}
\BIBentrySTDinterwordspacing

\bibitem{lu2021automatic}
J.~Lu, W.~Chen, H.~Ye, J.~Wang, H.~Mei, Y.~Gu, Y.~Wu, X.~L. Zhang, and K.-L. Ma, ``Automatic generation of unit visualization-based scrollytelling for impromptu data facts delivery,'' in \emph{2021 IEEE 14th Pacific visualization symposium (PacificVis)}.\hskip 1em plus 0.5em minus 0.4em\relax IEEE, 2021, pp. 21--30.

\bibitem{morth2022scrollyvis}
E.~M{\"o}rth, S.~Bruckner, and N.~N. Smit, ``Scrollyvis: Interactive visual authoring of guided dynamic narratives for scientific scrollytelling,'' \emph{IEEE Transactions on Visualization and Computer Graphics}, 2022.

\bibitem{pettersenscrollytelling}
M.~Pettersen, ``What is scrollytelling?(+ its impact on digital content): An interview.''

\bibitem{kosara_scrollytelling_2016}
\BIBentryALTinterwordspacing
R.~Kosara, ``\BIBforeignlanguage{en-US}{The {Scrollytelling} {Scourge}},'' May 2016. [Online]. Available: \url{https://eagereyes.org/blog/2016/the-scrollytelling-scourge}
\BIBentrySTDinterwordspacing

\bibitem{parker2019snowball}
C.~Parker, S.~Scott, and A.~Geddes, ``Snowball sampling,'' \emph{SAGE research methods foundations}, 2019.

\bibitem{shu2020makes}
X.~Shu, A.~Wu, J.~Tang, B.~Bach, Y.~Wu, and H.~Qu, ``What makes a data-gif understandable?'' \emph{IEEE Transactions on Visualization and Computer Graphics}, vol.~27, no.~2, pp. 1492--1502, 2020.

\bibitem{ge2020canis}
T.~Ge, Y.~Zhao, B.~Lee, D.~Ren, B.~Chen, and Y.~Wang, ``Canis: A high-level language for data-driven chart animations,'' in \emph{Computer Graphics Forum}, vol.~39, no.~3.\hskip 1em plus 0.5em minus 0.4em\relax Wiley Online Library, 2020, pp. 607--617.

\bibitem{wainer_1984}
\BIBentryALTinterwordspacing
H.~Wainer, ``How to display data badly,'' \emph{The American Statistician}, vol.~38, no.~2, pp. 137--147, 1984. [Online]. Available: \url{http://www.jstor.org/stable/2683253}
\BIBentrySTDinterwordspacing

\bibitem{kosslyn_1985}
\BIBentryALTinterwordspacing
S.~M. Kosslyn, ``Graphics and human information processing: A review of five books,'' \emph{Journal of the American Statistical Association}, vol.~80, no. 391, pp. 499--512, 1985. [Online]. Available: \url{http://www.jstor.org/stable/2288463}
\BIBentrySTDinterwordspacing

\bibitem{Zacks1998ReadingBG}
J.~M. Zacks, E.~Levy, B.~Tversky, and D.~J. Schiano, ``Reading bar graphs: Effects of extraneous depth cues and graphical context.'' \emph{Journal of Experimental Psychology: Applied}, vol.~4, pp. 119--138, 1998.

\bibitem{happiness_2021}
\BIBentryALTinterwordspacing
N.~Glozier and R.~Morris, ``Why happiness is becoming more expensive and out of reach,'' 2021. [Online]. Available: \url{https://www.sydney.edu.au/news-opinion/news/2021/11/16/why-happiness-is-becoming-more-expensive-and-out-of-reach-.html}
\BIBentrySTDinterwordspacing

\bibitem{Figueiras2014}
A.~Figueiras, ``How to tell stories using visualization,'' in \emph{2014 18th International Conference on Information Visualisation}, 2014, pp. 18--18.

\bibitem{hullman2013deeper}
J.~Hullman, S.~Drucker, N.~H. Riche, B.~Lee, D.~Fisher, and E.~Adar, ``A deeper understanding of sequence in narrative visualization,'' \emph{IEEE Transactions on visualization and computer graphics}, vol.~19, no.~12, pp. 2406--2415, 2013.

\bibitem{borkin2013makes}
M.~A. Borkin, A.~A. Vo, Z.~Bylinskii, P.~Isola, S.~Sunkavalli, A.~Oliva, and H.~Pfister, ``What makes a visualization memorable?'' \emph{IEEE transactions on visualization and computer graphics}, vol.~19, no.~12, pp. 2306--2315, 2013.

\bibitem{pandey2014persuasive}
A.~V. Pandey, A.~Manivannan, O.~Nov, M.~Satterthwaite, and E.~Bertini, ``The persuasive power of data visualization,'' \emph{IEEE transactions on visualization and computer graphics}, vol.~20, no.~12, pp. 2211--2220, 2014.

\bibitem{muehlenhaus2012if}
I.~Muehlenhaus, ``If looks could kill: The impact of different rhetorical styles on persuasive geocommunication,'' \emph{The Cartographic Journal}, vol.~49, no.~4, pp. 361--375, 2012.

\bibitem{liem2020structure}
J.~Liem, C.~Perin, and J.~Wood, ``Structure and empathy in visual data storytelling: Evaluating their influence on attitude,'' in \emph{Computer Graphics Forum}, vol.~39, no.~3.\hskip 1em plus 0.5em minus 0.4em\relax Wiley Online Library, 2020, pp. 277--289.

\bibitem{hidi2019interest}
S.~E. Hidi and K.~A. Renninger, ``Interest development and its relation to curiosity: Needed neuroscientific research,'' \emph{Educational Psychology Review}, vol.~31, no.~4, pp. 833--852, 2019.

\bibitem{Kukkonen2009}
H.~Oinas-Kukkonen and M.~Harjumaa, ``Persuasive systems design: Key issues, process model, and system features,'' \emph{Communications of the Association for Information Systems}, vol.~24, 03 2009.

\bibitem{shi2020}
D.~Shi, X.~Xu, F.~Sun, Y.~Shi, and N.~Cao, ``Calliope: Automatic visual data story generation from a spreadsheet,'' \emph{IEEE Transactions on Visualization and Computer Graphics}, vol.~PP, pp. 1--1, 10 2020.

\bibitem{skau2015evaluation}
D.~Skau, L.~Harrison, and R.~Kosara, ``An evaluation of the impact of visual embellishments in bar charts,'' in \emph{Computer Graphics Forum}, vol.~34, no.~3.\hskip 1em plus 0.5em minus 0.4em\relax Wiley Online Library, 2015, pp. 221--230.

\bibitem{Haroz_2015_isotype}
\BIBentryALTinterwordspacing
S.~Haroz, R.~Kosara, and S.~L. Franconeri, ``Isotype visualization: Working memory, performance, and engagement with pictographs,'' in \emph{Proceedings of the 33rd Annual ACM Conference on Human Factors in Computing Systems}, ser. CHI '15.\hskip 1em plus 0.5em minus 0.4em\relax New York, NY, USA: Association for Computing Machinery, 2015, p. 1191–1200. [Online]. Available: \url{https://doi.org/10.1145/2702123.2702275}
\BIBentrySTDinterwordspacing

\bibitem{morais2020showing}
L.~Morais, Y.~Jansen, N.~Andrade, and P.~Dragicevic, ``Showing data about people: A design space of anthropographics,'' \emph{IEEE Transactions on Visualization and Computer Graphics}, vol.~28, no.~3, pp. 1661--1679, 2020.

\bibitem{Jennett2008}
C.~Jennett, A.~L. Cox, P.~Cairns, S.~Dhoparee, A.~Epps, T.~Tijs, and A.~Walton, ``Measuring and defining the experience of immersion in games,'' \emph{International journal of human-computer studies}, vol.~66, no.~9, pp. 641--661, 2008.

\bibitem{stolper_emerging_2016}
C.~D. Stolper, B.~Lee, N.~Henry~Riche, and J.~Stasko, ``Emerging and {Recurring} {Data}-{Driven} {Storytelling} {Techniques}: {Analysis} of a {Curated} {Collection} of {Recent} {Stories},'' Tech. Rep. MSR-TR-2016-14, Apr. 2016.

\bibitem{ren2019visualization}
D.~Ren, ``Visualization authoring for data-driven storytelling,'' Ph.D. dissertation, UC Santa Barbara, 2019.

\bibitem{cordeil2017imaxes}
M.~Cordeil, A.~Cunningham, T.~Dwyer, B.~H. Thomas, and K.~Marriott, ``Imaxes: Immersive axes as embodied affordances for interactive multivariate data visualisation,'' in \emph{Proceedings of the 30th annual ACM symposium on user interface software and technology}, 2017, pp. 71--83.

\bibitem{sicat2018dxr}
R.~Sicat, J.~Li, J.~Choi, M.~Cordeil, W.-K. Jeong, B.~Bach, and H.~Pfister, ``Dxr: A toolkit for building immersive data visualizations,'' \emph{IEEE transactions on visualization and computer graphics}, vol.~25, no.~1, pp. 715--725, 2018.

\bibitem{evin2022cine}
I.~Evin, P.~H{\"a}m{\"a}l{\"a}inen, and C.~Guckelsberger, ``Cine-ai: Generating video game cutscenes in the style of human directors,'' \emph{Proceedings of the ACM on Human-Computer Interaction}, vol.~6, no. CHI PLAY, pp. 1--23, 2022.

\bibitem{huang2019learning}
C.~Huang, Z.~Yang, Y.~Kong, P.~Chen, X.~Yang, and K.-T.~T. Cheng, ``Learning to capture a film-look video with a camera drone,'' in \emph{2019 international conference on robotics and automation (ICRA)}.\hskip 1em plus 0.5em minus 0.4em\relax IEEE, 2019, pp. 1871--1877.

\bibitem{he2023virtual}
L.-w. He, M.~F. Cohen, and D.~H. Salesin, ``The virtual cinematographer: A paradigm for automatic real-time camera control and directing,'' in \emph{Seminal Graphics Papers: Pushing the Boundaries, Volume 2}, 2023, pp. 707--714.

\end{thebibliography}

\subsection{Appendix: Data Story Links}
\begin{enumerate}
\renewcommand{\labelenumi}{\theenumi}
\renewcommand{\theenumi}{[S\arabic{enumi}]}

\item\label{link:S1} \href{https://www.nytimes.com/interactive/2018/05/09/nyregion/subway-crisis-mta-decisions-signals-rules.html}{Pearce, A. (2018). How 2 M.T.A. Decisions Pushed the Subway Into Crisis. \emph{The New York Times.}}
\item\label{link:S2} \href{https://www.nytimes.com/interactive/2021/09/15/nyregion/empire-state-building-reopening-new-york.html}{Collins, K., Diamant, N., Eavis, P., Fleisher, O., Haag, M., Harvey, B., Huang, L., Patanjali, K., Peyton, M., \& Taylor, R. (2021). Why the Empire State Building, and New York, May Never Be the Same. \emph{The New York Times.}}
\item\label{link:S3} \href{https://buildinghop.es/}{Accurat. (2018.) Google Building Hopes} .
\item\label{link:S4} \href{https://pudding.cool/2017/10/satellites/}{Conlen, M. (2017). Seeing Earth from Outer Space. \emph{The Pudding.}}
\item\label{link:S5} \href{https://pudding.cool/2018/12/3d-cities-story/}{Daniels, M. (2018). Population Mountains. \emph{The Pudding}.}
\item\label{link:S6} \href{https://www.nytimes.com/interactive/2015/03/19/upshot/3d-yield-curve-economic-growth.html}{Aisch, G., \& Cox, A. (2015). A 3-D View of a Chart That Predicts The Economic Future: The Yield Curve. \emph{The New York Times,}.}
\item\label{link:S7} \href{https://www.nytimes.com/interactive/2021/01/28/opinion/climate-change-risks-by-country.html}{Serkez, Y. (2021). Every Country Has Its Own Climate Risks. What’s Yours? \emph{The New York Times,}.}
\item\label{link:S8} \href{https://pudding.cool/2018/10/city_3d/}{Daniels, M. (2018). Human Terrain. \emph{The Pudding.}}
\item\label{link:S9} \href{https://pudding.cool/2018/03/neighborhoods/}{Blinderman, I., \& Thomas, A. (2018). A Tale of Two Cities. \emph{The Pudding.}}
\item\label{link:S10} \href{https://www.nytimes.com/interactive/2021/01/14/climate/hottest-year-2020-global-map.html}{Fountain, H., Migliozzi, B., \& Popovich, N. (2021). Where 2020's Record Heat Was Felt the Most. \emph{The New York Times.}}
\item\label{link:S11} \href{https://www.nytimes.com/interactive/2021/05/24/us/tulsa-race-massacre.html}{Parshina-Kottas, Y., Singhvi, A., Burch, A., Griggs, T., Gröndahl, M., Huang, L., Wallace, T., White, J., \& Williams, J. (2021). What The Tulsa Race Massacre Destroyed. \emph{The New York Times.}}
\item\label{link:S12} \href{https://www.nytimes.com/interactive/2021/06/30/opinion/environmental-inequity-trees-critical-infrastructure.html}{Leahy, I., \& Serkez, Y. (2021). Since When Have Trees Existed Only for Rich Americans? \emph{The New York Times.}}
\item\label{link:S13} \href{https://pudding.cool/2018/07/women-in-congress/}{D'souza, D. (2018). We mapped out the road to gender parity in the House of Representatives \emph{The Pudding.}}
\item\label{link:S14} \href{https://pudding.cool/2018/02/stand-up/}{Samora, R., \& Daniels, M. (2018). Structure of Stand Up Comedy. \emph{The Pudding}}
\item\label{link:S15} \href{https://datanibbl.es/tracing-kpop-wave/}{Tracing the K-POP WAVE. (n.d.).}
\item\label{link:S16} \href{https://pudding.cool/2017/01/making-it-big/}{Samora, R., \& Kopf, D. (2017). The Unlikely Odds of Making It Big. \emph{The Pudding.}}
\item\label{link:S17} \href{https://www.nytimes.com/interactive/2021/03/10/opinion/covid-vaccine-lines-states.html}{Lash, N. (2021).  Who’s Next in Your State’s Vaccine Line? \emph{The New York Times.}}
\item\label{link:S18} \href{https://www.nytimes.com/interactive/2021/07/30/sports/olympics/olympic-running.html}{Buchanan, L., Kessel, J.M., Rhyne, E., Throop, N., Ward, J., \& White, J. (2021). How Speed and Distance Dictate How Olympians Run. \emph{The New York Times.}}
\item\label{link:S19} \href{https://www.theatlantic.com/theplatinumpatients/}{McGill, A., \& Goldenberg, R. (2017). The Platinum Patients. \emph{The Atlantic.}}
\item\label{link:S20} \href{https://pudding.cool/2017/04/beer/}{Samora, R. (2017). Craft beer — so hot right now. But what city is the microbrew capital of the US? \emph{The Pudding.}}
\item\label{link:S21} \href{https://pudding.cool/2021/03/love-and-ai/}{Mishkin, P., Samora, R., \& Diehm, J. (2021). Nothing Breaks Like A.I. Heart. \emph{The Pudding.}}
\item\label{link:S22} \href{https://pudding.cool/2017/03/punk/}{Daniels, M. (2017). Crowdsourcing the Definition of "Punk". \emph{The Pudding.}}
\item\label{link:S23} \href{https://pudding.cool/2018/08/retraining/}{Dworkin, J., \& Blinderman, I. (2018). Why the tech sector may not solve America’s looming automation crisis. \emph{The Pudding.}}

\end{enumerate}

\subsection{Biographies and Author Photos}
\vskip -3\baselineskip plus -1fil
\begin{IEEEbiography}[{\includegraphics[clip,trim={1.5cm 0cm 1.5cm 0cm},width=1in,height=1.25in]{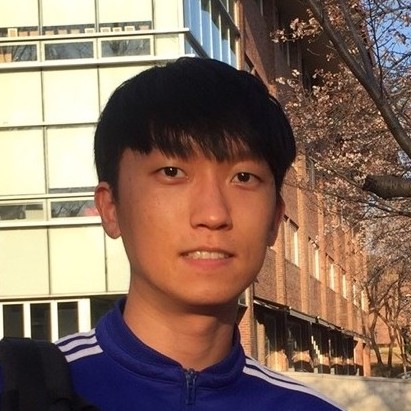}}]
{Seon Gyeom Kim} is a Ph.D. candidate in Industrial Design Department at KAIST. 
His research encompasses human-computer interaction, data visualization and AI, focusing on integrating large language model with data story creation.
\end{IEEEbiography}

\vskip -1\baselineskip plus -1fil
\begin{IEEEbiography}[{\includegraphics[clip,trim={5cm 0 5cm 0},width=1in,height=1.25in]{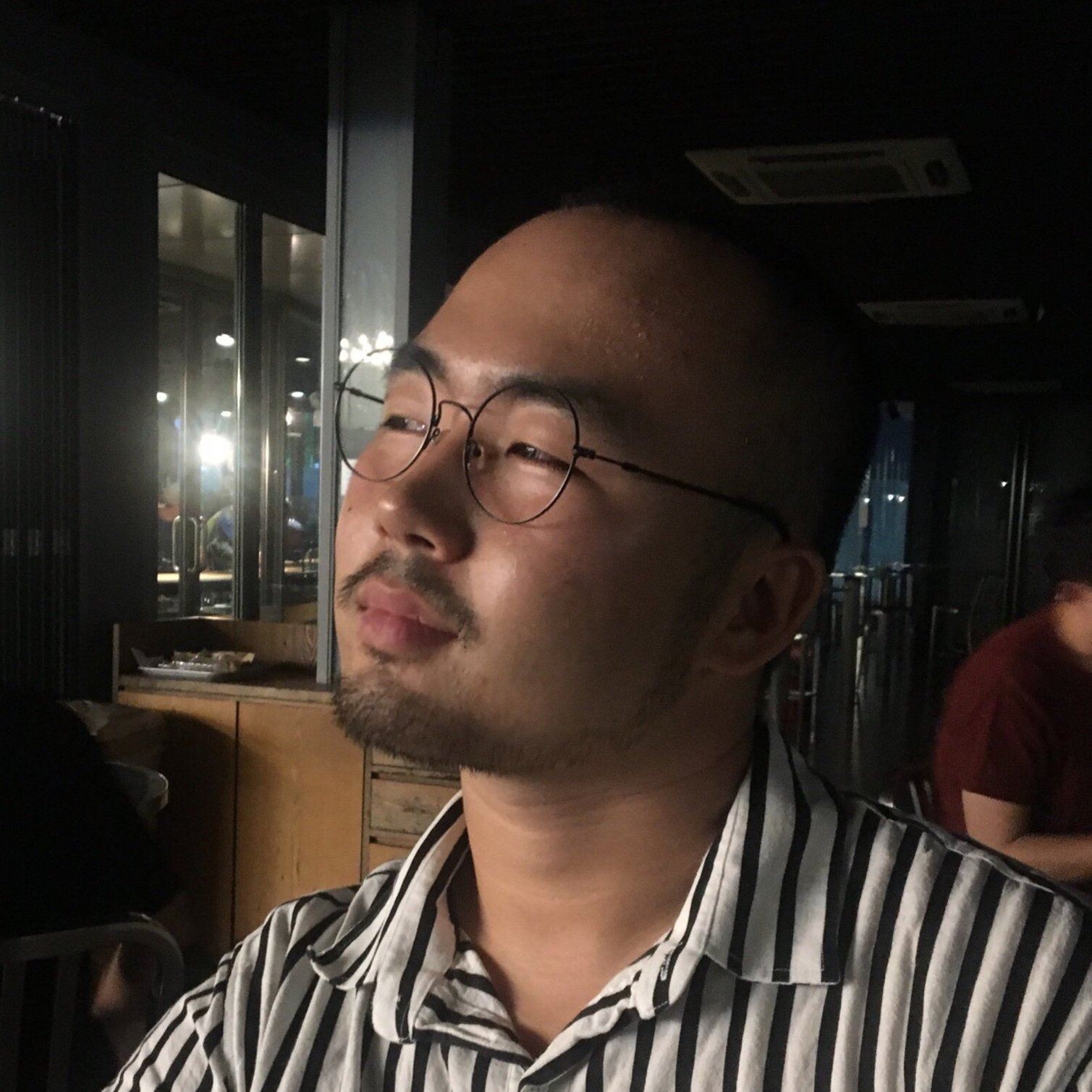}}]
{Juhyeong Park} has a master's degree in Industrial Design from KAIST, with a focus on human-computer interaction. He is committed to leveraging AI-based design approaches to create impactful solutions.
\end{IEEEbiography}

\vskip -1\baselineskip plus -1fil
\begin{IEEEbiographynophoto}{Yutaek Song} is a researcher in NAVER. He graduated from KAIST with a bachelor's degree in Industrial Design.
\end{IEEEbiographynophoto}

\vskip -1\baselineskip plus -1fil
\begin{IEEEbiography}[{\includegraphics[clip,trim={0 0 0 0},width=1in,height=1.25in]{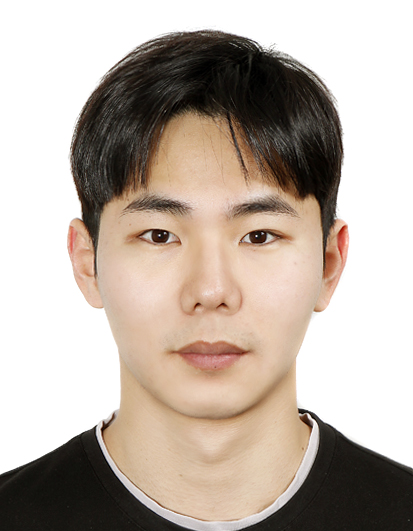}}]
{Donggun Lee} is an undergraduate student at KAIST, majoring in Industrial Design with a minor in AI. His research interests lie in the field of Human-Computer Interaction (HCI), with a particular focus on Human-AI Interaction (HAI). He is committed to exploring these interactions further, aiming to utilize data-driven design approaches to create impactful solutions.
\end{IEEEbiography}

\vskip -1\baselineskip plus -1fil
\begin{IEEEbiographynophoto}{Yubin Lee} is a master student at the Computer Science Department of KAIST. She earned her Bachelor's degree in Industrial Design at KAIST.
\end{IEEEbiographynophoto}

\vskip -1\baselineskip plus -1fil
\begin{IEEEbiography}[{\includegraphics[clip,trim={0 0 0.3cm 0},width=1in,height=1.25in]{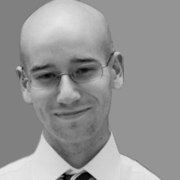}}]
{Ryan A. Rossi} is a machine learning research scientist at Adobe Research in San Jose, CA. His research lies in the fields of machine learning; and spans theory, algorithms, and applications of large complex relational (network/graph) data from social and physical phenomena. 
Ryan earned his Ph.D. and M.S. in Computer Science at Purdue University. 
\end{IEEEbiography}

\vskip -1\baselineskip plus -1fil
\begin{IEEEbiography}[{\includegraphics[clip,trim={3cm 0 1cm 0},width=1in,height=1.25in]{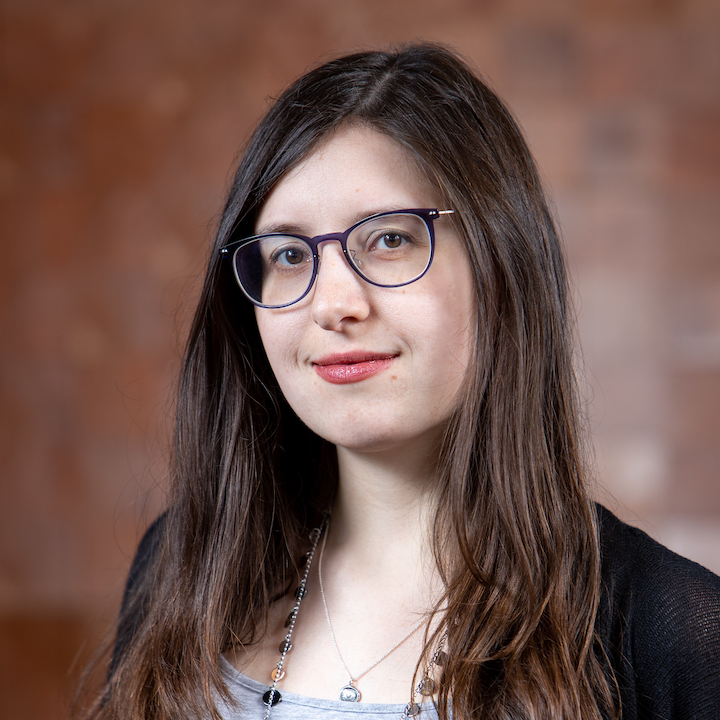}}]
{Jane Hoffswell} has a Ph.D. in Computer Science from the University of Washington (2020). Jane is currently a research scientist at Adobe, specializing in visualization and human-computer interaction. Her research generally explores the design of responsive visualizations, interactive systems, and visual storytelling.
\end{IEEEbiography}

\vskip -1\baselineskip plus -1fil
\begin{IEEEbiography}[{\includegraphics[clip,trim={5.5cm 0 5.5cm 0},width=1in,height=1.25in]{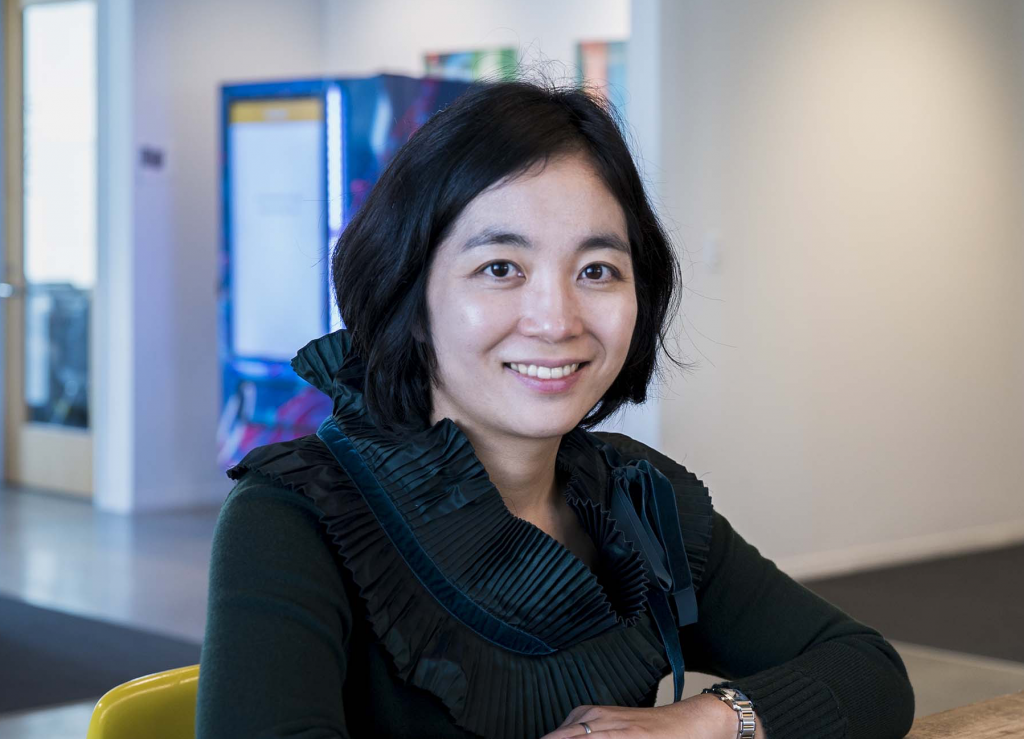}}]
{Eunyee} is a Principal Scientist at Adobe Research. Her research interests lies at the intersection of Human Computer Interaction, AI/Machine Learning, and Data Science. She is passionate about how to tailor and personalize information for people, then they can access the info they want much more efficiently. She finished her PhD at the Computer Science Department of Texas A\&M University. Her advisor is Prof. Andruid Kerne. 
She graduated in 2002 from the Seoul National University with a bachelor’s in computer science and engineering.
\end{IEEEbiography}

\vskip -1\baselineskip plus -1fil
\begin{IEEEbiography}[{\includegraphics[clip,trim={33cm 40cm 34cm 16cm},width=1in,height=1.25in]{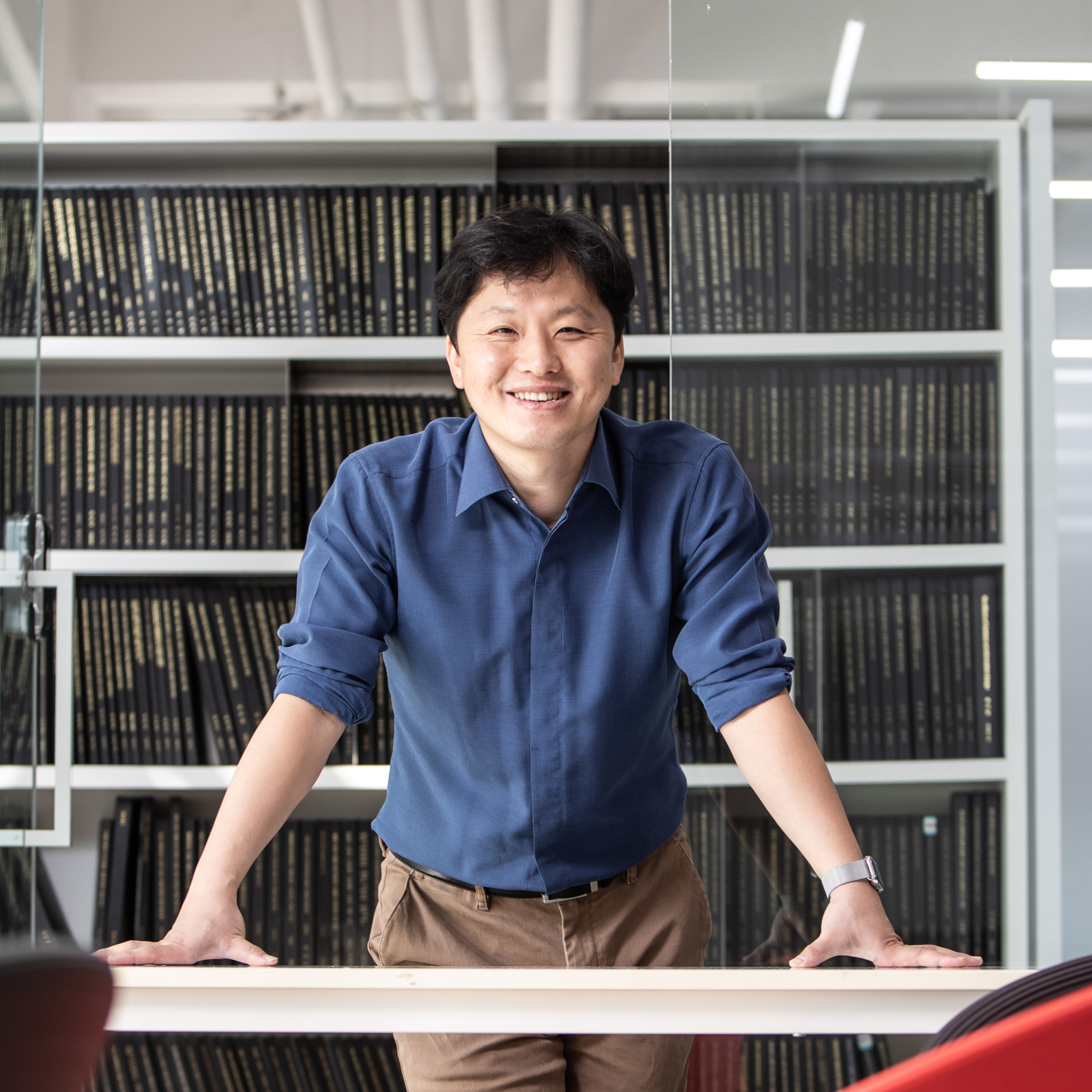}}]
{Tak Yeon Lee} is an assistant professor at KAIST, Daejeon, Republic of Korea. He received his PhD in Computer Science from University of Maryland, College Park, USA. His research focuses on promoting symbiotic collaboration between humans and AI.
\end{IEEEbiography}

\end{document}